\documentclass[twocolumn]{aastex701}

\usepackage{amsmath}
\usepackage{longtable}
\usepackage{multirow}

\newcommand{\pa} {\mathrm{P.A.}}

\newcommand{\h} {^{\mathrm{h}}}
\newcommand{\m} {^{\mathrm{m}}}
\newcommand{\s} {^{\mathrm{s}}}

\newcommand{\mJybeam}  {\mbox{mJy}~\mbox{beam}^{-1}}

\newcommand{\kms}	{\mbox{km s}^{-1}}

\newcommand{\K}	{{\rm K}}

\shorttitle{Hot Disk}
\shortauthors{Yang et al.}

\begin{document}

\title{HOTDISK. Finding Massive Protostellar Disks with Water and Refractory Molecular Species}

\author[orcid=0000-0002-9839-185X]{Kai Yang}
\affiliation{State Key Laboratory of Dark Matter Physics, School of Physics and Astronomy, Shanghai Jiao Tong University, Shanghai 200240, People’s Republic of China}
\affiliation{Department of Astronomy, School of Physics and Astronomy, Shanghai Jiao Tong University, 800 Dongchuan Road, Shanghai 200240, People’s Republic of China}
\affiliation{Key Laboratory for Particle Astrophysics and Cosmology (MOE) / Shanghai Key Laboratory for Particle Physics and Cosmology, Shanghai 200240, People’s Republic of China}
\email[show]{kyang2146@sjtu.edu.cn}  

\author[orcid=0000-0001-7511-0034]{Yichen Zhang}
\affiliation{State Key Laboratory of Dark Matter Physics, School of Physics and Astronomy, Shanghai Jiao Tong University, Shanghai 200240, People’s Republic of China}
\affiliation{Department of Astronomy, School of Physics and Astronomy, Shanghai Jiao Tong University, 800 Dongchuan Road, Shanghai 200240, People’s Republic of China}
\affiliation{Key Laboratory for Particle Astrophysics and Cosmology (MOE) / Shanghai Key Laboratory for Particle Physics and Cosmology, Shanghai 200240, People’s Republic of China}
\email[show]{yczhang.astro@gmail.com}  

\author[orcid=0000-0002-6907-0926]{Kei E. I. Tanaka}
\affiliation{Department of Earth and Planetary Sciences, Institute of Science Tokyo, Meguro, Tokyo 152-8551, Japan}
\email{kt503i@gmail.com}  

\author[orcid=0000-0002-5286-2564]{Tie Liu}
\affiliation{State Key Laboratory of Radio Astronomy and Technology, Shanghai Astronomical Observatory, Chinese Academy of Sciences, \\
80 Nandan Road, Shanghai 200030, People's Republic of China}
\email{liutie@shao.ac.cn}

\author[orcid=0000-0002-3297-4497]{Nami Sakai}
\affiliation{RIKEN Cluster for Pioneering Research, 2-1, Hirosawa, Wako-shi, Saitama 351-0198, Japan}
\email{nami.sakai@riken.jp}

\author[orcid=0000-0002-9927-2705]{Ziwei E. Zhang}
\affiliation{RIKEN Cluster for Pioneering Research, 2-1, Hirosawa, Wako-shi, Saitama 351-0198, Japan}
\email{ziwei.zhang@riken.jp}

\author[orcid=0009-0002-9273-7220]{Gyuho Lee}
\affiliation{Korea Astronomy and Space Science Institute, 776 Daedeokdae-ro, Yuseong-gu, Daejeon 34055, Republic of Korea}
\affiliation{University of Science and Technology (UST), 217 Gajeong-ro, Yuseong-gu, Daejeon 34113, Republic of Korea}
\email{space295@kasi.re.kr}

\author[orcid=0000-0003-2412-7092]{Kee-Tae Kim}
\affiliation{Korea Astronomy and Space Science Institute, 776 Daedeokdae-ro, Yuseong-gu, Daejeon 34055, Republic of Korea}
\affiliation{University of Science and Technology (UST), 217 Gajeong-ro, Yuseong-gu, Daejeon 34113, Republic of Korea}
\email{ktkim@kasi.re.kr}

\author[orcid=0000-0001-6431-9633]{Adam Ginsburg}
\affiliation{Department of Astronomy, University of Florida, P.O.\ Box 112055, Gainesville, FL 32611, USA}
\email{adam.g.ginsburg@gmail.com}

\author[orcid=0000-0002-6540-7042]{Lile Wang}
\affiliation{Kavli Institute for Astronomy and Astrophysics, Peking University, Beijing 100871, People’s Republic of China}
\affiliation{Department of Astronomy, School of Physics, Peking University, Beijing 100871, People’s Republic of China}
\email{lilew@pku.edu.cn}

\author[orcid=0009-0007-9664-0270]{Yao Wang}
\affiliation{State Key Laboratory of Dark Matter Physics, School of Physics and Astronomy, Shanghai Jiao Tong University, Shanghai 200240, People’s Republic of China}
\affiliation{Department of Astronomy, School of Physics and Astronomy, Shanghai Jiao Tong University, 800 Dongchuan Road, Shanghai 200240, People’s Republic of China}
\affiliation{Key Laboratory for Particle Astrophysics and Cosmology (MOE) / Shanghai Key Laboratory for Particle Physics and Cosmology, Shanghai 200240, People’s Republic of China}
\email{sjtu1687090@sjtu.edu.cn}

\author[orcid=0009-0004-9907-3716]{Yongzhi Tang}
\affiliation{State Key Laboratory of Dark Matter Physics, School of Physics and Astronomy, Shanghai Jiao Tong University, Shanghai 200240, People’s Republic of China}
\affiliation{Department of Astronomy, School of Physics and Astronomy, Shanghai Jiao Tong University, 800 Dongchuan Road, Shanghai 200240, People’s Republic of China}
\affiliation{Key Laboratory for Particle Astrophysics and Cosmology (MOE) / Shanghai Key Laboratory for Particle Physics and Cosmology, Shanghai 200240, People’s Republic of China}
\email{tangyongzhi@sjtu.edu.cn}

\author[orcid=0000-0002-8691-4588]{Yu Cheng}
\affiliation{National Astronomical Observatory of Japan, 2-21-1 Osawa, Mitaka, Tokyo 181-8588, Japan}
\affiliation{Deptartment of Astronomy, University of Virginia, Charlottesville, VA 22904, USA}
\email{ycheng.astro@gmail.com}

\author[orcid=0000-0003-3343-9645]{Hongli Liu}
\affiliation{School of Physics and Astronomy, Yunnan University, Kunming 650091, People’s Republic of China}
\email{hongliliu2012@gmail.com}

\author[orcid=0000-0001-9822-7817]{Wenyu Jiao}
\affiliation{State Key Laboratory of Radio Astronomy and Technology, Shanghai Astronomical Observatory, Chinese Academy of Sciences, \\
80 Nandan Road, Shanghai 200030, People's Republic of China}
\email{astrojiao@gmail.com}

\author[orcid=0000-0001-5950-1932]{Fengwei Xu}
\affiliation{Max Planck Institute for Astronomy, Königstuhl 17, 69117 Heidelberg, Germany}
\email{fengwei@mpia.de}

\author[orcid=0000-0001-8315-4248]{Xunchuan Liu}
\affiliation{State Key Laboratory of Radio Astronomy and Technology, Shanghai Astronomical Observatory, Chinese Academy of Sciences, \\
80 Nandan Road, Shanghai 200030, People's Republic of China}
\email{liuxunchuan001@gmail.com}

\author[orcid=0000-0001-7573-0145]{Xiaofeng Mai}
\affiliation{State Key Laboratory of Radio Astronomy and Technology, Shanghai Astronomical Observatory, Chinese Academy of Sciences, \\
80 Nandan Road, Shanghai 200030, People's Republic of China}
\email{maixf@shao.ac.cn}

\author[orcid=0009-0004-6159-5375]{Dongting Yang}
\affiliation{School of Physics and Astronomy, Yunnan University, Kunming 650091, People’s Republic of China}
\email{dongting@mail.ynu.edu.cn}

\correspondingauthor{Kai Yang, Yichen Zhang}


\begin{abstract}
We present high-angular-resolution ($\sim0.05^{\prime\prime}$, $\sim 60-250$ au)
ALMA Band~6 observations from the HOTDISK project 
(Hot-Origin Tracer survey of DISKs of massive protostars) 
aimed at investigating the ``hot-disk'' chemical pattern traced by 
vibrationally excited water, NaCl, SiS, and SiO in the innermost regions 
around massive protostars. 
Ten targets were selected based on strong CH$_3$CN emission exhibiting clear rotational signatures and centrally concentrated SiO emission from lower-resolution observations.
We detect vibrationally excited water emission toward 7 of the 10 sources. 
In all detections, the blueshifted and redshifted components are compact and 
located on opposite sides of the 1.3 mm continuum peak, 
with velocity gradients approximately perpendicular to the outflow axes, 
consistent with rotation on disk scales. 
Emission from NaCl and SiS is detected toward 5 of these 7 sources and 
exhibits similar kinematics, 
further supporting the presence of compact rotating structures.
In contrast, commonly used hot-core tracers (e.g., CH$_3$CN and SO$_2$) 
primarily probe larger-scale envelope gas. 
These results demonstrate that vibrationally excited water, NaCl, and SiS 
are powerful tracers of disk structures on $\sim$100 au scales, 
when observed at sufficient angular resolution and sensitivity.
The high detection rate suggests that hot-disk chemical patterns -- 
and thus compact rotating disks -- 
are common in massive star-forming regions, 
at least among sources with well-developed rotating envelopes.
\end{abstract}




\section{Introduction}
\label{sec:intro}

Massive stars play a fundamental role across many areas of astrophysics, 
yet their formation remains poorly understood. 
A central question is whether massive protostars accrete material through 
rotationally supported disks, 
analogous to those observed around low-mass protostars. 
From a theoretical perspective, disk accretion is expected, 
as disks provide an efficient mechanism to channel material onto the central object 
while redistributing angular momentum mitigating the strong radiative feedback from the forming star 
\citep{2009Sci...323..754K,Tanaka2017_impact,2018A&A...616A.101K,2019ApJ...887..108R}. 
Such feedback was historically considered a major barrier to the formation of high-mass stars \citep{1977A&A....54..183Y,1987ApJ...319..850W}.

Despite this progress, our understanding of disk accretion 
in massive star formation remains incomplete. 
The current sample of massive protostellar disks is still small, 
largely due to the observational challenges posed by their large distances 
and the clustered environments in which they form. 
Moreover, while velocity gradients consistent with differential rotation 
are frequently observed, these signatures often arise in massive ($\sim$10$-$100 $M_{\odot}$) 
and extended (10$^{3}$–10$^{4}$ au) structures 
-- commonly referred to as envelopes or ``toroids'' \citep{2007prpl.conf..197C,2014prpl.conf..149T,2016A&ARv..24....6B,2017A&A...602A..59C} -- 
rather than in compact, rotationally supported disks. 
In many cases, the presence of a true Keplerian disk remains ambiguous, 
as the contributions from disks and rotating envelopes are not clearly disentangled 
\citep[e.g.,][]{2026ApJ...999..106O}.
This difficulty is compounded by the fact that commonly used molecular tracers, 
such as CH$_3$CN, may preferentially trace warm envelope material 
rather than disk gas, since in massive protostars the envelope can be warm enough for complex organic molecules, primarily formed on dust grain surfaces, to desorb into the gas-phase \citep{2018A&A...620A..31M,Zhang2019massive,Tanaka2020}. 
These limitations highlight the need for observations 
employing a broader range of molecular tracers 
to robustly identify and characterize disks.

Recent studies have identified a distinctive chemical signature 
in some massive protostellar disks, 
referred to as the “hot-disk” chemical pattern \citep{Tanaka2020}.
This pattern is characterized by emission from gaseous species 
of alkali halides (e.g., NaCl and KCl) and silicon-bearing molecules (e.g., SiO and SiS). 
In extreme cases, refractory metal-bearing species such as AlO have also been detected 
\citep[e.g.,][]{2019ApJ...875L..29T}.
These molecules are thought to enter the gas phase as refractory dust grain cores 
are processed in the inner regions, 
although it remains unclear whether this occurs primarily through thermal desorption 
or through the destruction of dust grains. 
In this paper, we use the term ``refractory species'' in a broad sense 
to refer to molecules 
which are preferentially associated with regions close to the protostar 
and likely arise from high-temperature processing of refractory dust,
including NaCl and SiS,
although their detailed formation mechanisms remain uncertain.
In addition, vibrationally excited H$_2$O lines are often observed and 
are closely associated with these refractory species in both spatial distribution and kinematics \citep[e.g.,][]{Tanaka2020,Ginsburg2023}. 

One of the best-studied examples is Orion Source I 
\citep{2017NatAs...1E.146H,2019ApJ...872...54G,2019ApJ...875L..29T,Ginsburg2023},
with additional detections reported toward sources such as 
I16547$-$4247 \citep{Tanaka2020} 
and G17.64+0.16 \citep{2019A&A...627L...6M,Ginsburg2023}. 
In these systems, ``hot-disk'' molecules are found to trace only 
the inner $\sim$100–200 au regions, consistent with disk scales, 
while more volatile species (e.g., SO$_2$ and CH$_3$CN) 
predominantly trace the surrounding envelope. 
A systematic search of 17 well-known massive young stellar objects (MYSOs) 
using ALMA archival data has identified six MYSOs
-- including the three sources mentioned above -- 
exhibiting this “hot-disk” chemical pattern, 
corresponding to a total of nine sources when accounting for 
binary and multiple systems \citep{Ginsburg2023}.

Although the association of ``hot-disk'' species with high-temperature dust processing
in the inner disk has been suggested,
the physical and chemical conditions required to produce such emission 
remain poorly constrained. 
Nevertheless, the “hot-disk” chemical pattern offers a 
promising diagnostic for identifying and probing massive protostellar disks. 
If these lines indeed exclusively trace disk material, 
they can provide direct and unambiguous evidence for the presence of disks.

In this work, as part of the 
Hot-Origin Tracer survey of DISKs of massive protostars (HOTDISK) project, 
we search for hot-disk signatures in massive star-forming regions 
using sensitive, high-angular-resolution ($\sim$0.05$^{\prime\prime}$) 
ALMA Band 6 observations toward a sample of ten MYSOs. 
The sample selection and observational setup are described in Section~\ref{sec:sample_obs}. 
The results and data analysis are presented in Section~\ref{sec:results}. 
In Section~\ref{sec:dis}, we discuss the implications of our findings, 
and we summarize our conclusions in Section~\ref{sec:sum}.

\section{Sample selection and observations}
\label{sec:sample_obs}

Our pilot HOTDISK sample of 10 sources (listed in Table~\ref{tab:info}) 
was drawn from the ALMA-QUARKS survey \citep{Liu2024}, 
which observed 139 massive star-forming regions 
at an angular resolution of $0.3^{\prime\prime}$. 
From this parent sample, targets were selected based on the following criteria.
(1) Strong CH$_3$CN emission exhibiting clear rotational signatures 
on scales of several $\times 10^3$ au. 
Although CH$_3$CN primarily traces the envelope at $0.3^{\prime\prime}$ resolution, 
the presence of ordered rotation indicates a coherent rotating structure, 
within which a disk is likely present.
(2) Relatively bright and centrally concentrated SiO emission in the 
$0.3^{\prime\prime}$-resolution data. 
We identify centrally concentrated SiO emission 
from the moment 0 maps as emission whose peak lies within one synthesized beam 
of the continuum peak and whose intensity is significantly higher 
than that of the surrounding extended SiO emission.
While SiO is not universally detected in hot disks \citep{Ginsburg2023}, 
it is generally brighter than other refractory species 
and can remain detectable at moderate resolution. 
While SiO can also trace jets and outflows, limiting the SiO emission to cases 
showing a compact peak at the continuum position is essential 
to increase the likelihood that the detected signals are 
associated with the disk (or the base of a disk wind), 
rather than with shocked outflow regions.
In some cases, the SiO kinematics are also partially consistent 
with the CH$_3$CN velocity structure at this resolution.
Under these conditions, sources with bright, centrally concentrated SiO emission are 
therefore promising candidates for exhibiting the “hot-disk” chemical pattern. 
(3) Distances $\lesssim5$ kpc, ensuring sufficient physical resolution 
($\lesssim$250 au for a $0.05^{\prime\prime}$ beam) to probe disk-scale structures.

The observations were carried out with the ALMA 12 m array in Band 6 
(Program: 2024.1.01198.S, PI: K. Tanaka). 
The observational setup (including spectral configuration and array configurations) 
is identical to that used by \citet{Tanaka2020} for observations of I16547$-$4247. 
Although this source is part of the QUARKS sample and satisfies our selection criteria, 
it has already been established as a hot-disk source
and is therefore not included in our new observations.

For each source, data were obtained in two configurations: 
a more extended configuration (TM1; C-7/C-8) 
and a more compact configuration (TM2; C-4/C-5). 
The typical on-source integration times are $\sim$48 min for TM1 
and $\sim$10 min for TM2. 
Each target was observed with a single pointing, 
yielding a primary beam (half-power beam width) of $22^{\prime\prime}$. 
Details of the observations are summarized in Table~\ref{tab:obs}.

The correlator setup consists of four spectral windows, 
each with 1920 channels and a bandwidth of 1875 MHz,
covering rest-frequency ranges of 
216.66$-$218.54 GHz, 
219.26$-$221.14 GHz, 
231.46$-$233.34 GHz, 
and 233.56$-$235.44 GHz.
The channel width is 976.652 kHz, 
corresponding to 1.26 km s$^{-1}$ at 232.6867 GHz
(the rest frequency of H$_{2}$O $v_{2}=1$ $5_{5,0}-6_{4,3}$ transition), 

The data were calibrated using the CASA pipeline \citep[version 6.6.1,][]{mcmullin2007,2022PASP..134k4501C}. 
Following pipeline calibration, we performed self-calibration on the continuum 
constructed from line-free channels. 
This involved multiple iterations of phase-only self-calibration, 
with solution intervals decreasing from 60 s 
to the shortest values permitted by the data quality, 
followed by one or more iterations of amplitude self-calibration 
with solution intervals equal to the scan length, 
until no further improvement was achieved.

Self-calibration was first performed separately for each configuration, 
followed by a final combined self-calibration using scan-length solution intervals 
to align the datasets. 
The resulting calibration solutions were then applied to the spectral-line data. 
Imaging was carried out using the CASA {\it tclean} task with Briggs weighting 
and a fiducial robust parameter of 0.5. 
The low- and high-resolution continuum images 
have typical angular resolutions of 0.35$^{\prime\prime}$ and 0.05$^{\prime\prime}$, and rms noise levels of 60 and 30 $\mu$Jy beam$^{-1}$, respectively.
The high-resolution spectral line cubes typically have resolutions of 0.05$^{\prime\prime}$ and rms noise level of 0.6 mJy beam$^{-1}$ per channel.
For I18117$-$1753, a robust parameter of $-0.5$ is used for the high-resolution continuum image to better resolve the compact central structure. 
The resulting image has a rms noise level of 15 $\mu$Jy beam$^{-1}$.
The synthesized beam sizes and rms noise levels of each source
are listed in Table~\ref{tab:data}. 

\section{Results}
\label{sec:results}

\subsection{1.3 mm Continuum}
\label{sec:continuum}

\begin{figure*}[!ht]
\centering
\includegraphics[width=0.92\textwidth]{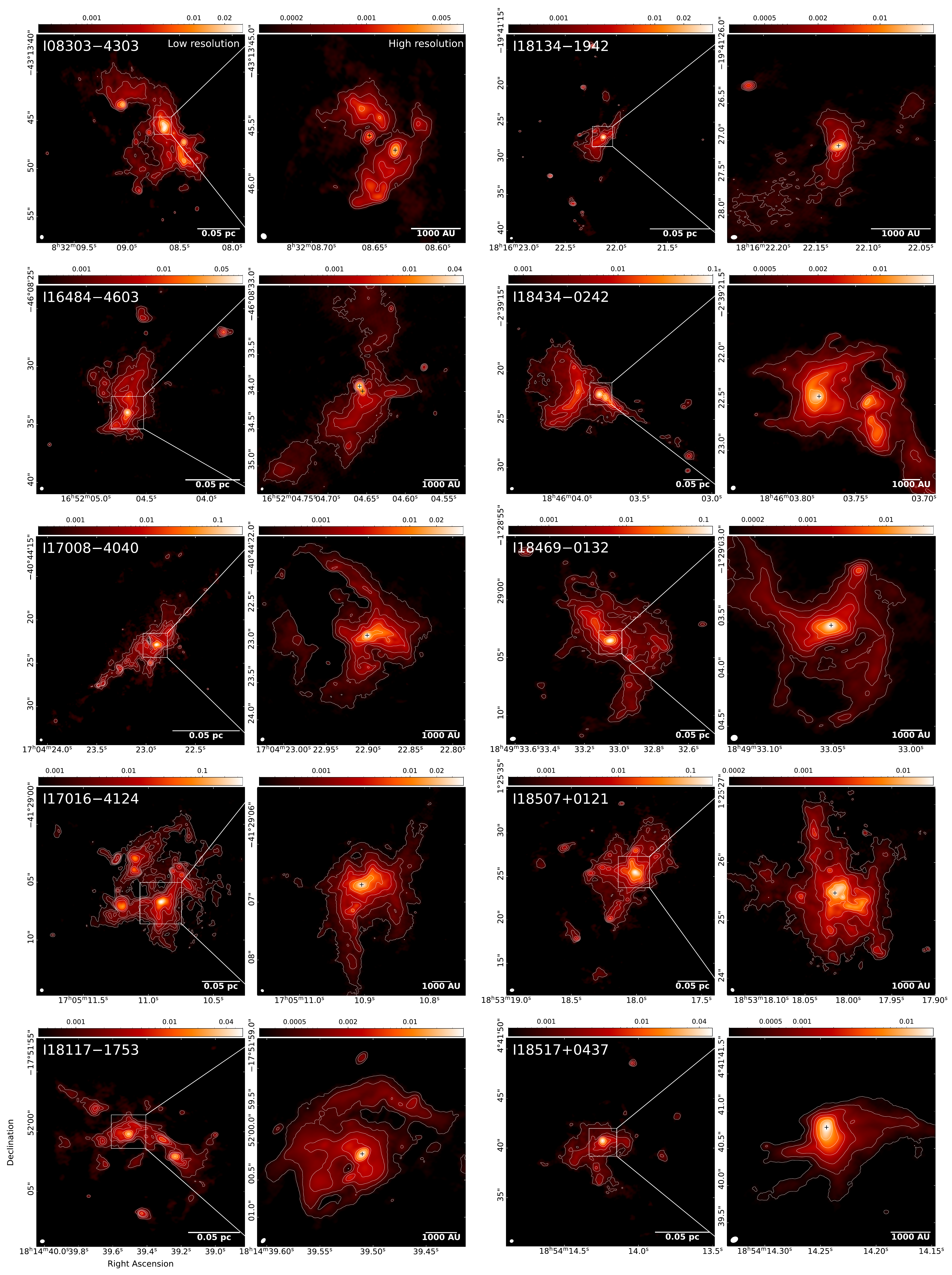}
\caption{ALMA 1.3 mm continuum images of the ten selected sources 
at low resolution (TM2) and high resolution (TM1+TM2). 
Contours are drawn at levels of $10\sigma \times 2^{n}$ ($n = 0, 1, 2, 3, \ldots$), 
where the rms noise levels $\sigma$ are listed in Table~\ref{tab:data}. 
The synthesized beam is shown in the bottom-left corner of each panel, 
and the scale bar is indicated in the bottom-right corner.
The crosses mark the continuum sources, including the continuum peaks and, in some cases, the centers of disk candidates.
The peak intensities of all continuum images are presented in Table \ref{tab:data}. The color bars show the intensity in units of Jy beam$^{-1}$.}
\label{fig:sample_conti}
\end{figure*}

Figure~\ref{fig:sample_conti} presents the 1.3 mm continuum emission 
of the ten sources observed with ALMA. 
For each target, we show two maps: 
a lower-resolution image obtained using the TM2 configuration 
and a higher-resolution image produced by combining the TM1 and TM2 data. 
The low-resolution maps present emission on scales of $\sim$0.2 pc, 
while the high-resolution images zoom into the 
central $\lesssim$ several $\times 10^{3}$ au region around the main source, 
where disk-scale structures can be better resolved. 
The comparison between the TM2-only and TM1+TM2 images provides a direct link 
between the core-scale emission and the disk and envelope structures 
within the inner few hundred au.

At low angular resolution, all sources exhibit bright, centrally concentrated 
continuum emission embedded within more extended structures. 
Several objects
(e.g., I08303$-$4303, I17008$-$4040, I18117$-$1753) 
show asymmetric or filamentary morphologies, 
suggesting that the central cores are not isolated but are connected 
to larger-scale dense material. 
The high-resolution maps resolve the central emission into compact components 
within the inner few thousand au.

In several sources (e.g., I08303$-$4303, I16484$-$4603, I17016$-$4124, and I18507+0121), 
the continuum emission fragments into multiple peaks within the central few hundred au, 
indicating the presence of close multiple systems or substructure within the inner core.
Other targets, such as I17008$-$4040, I18134$-$1942, and I18517+0437, 
exhibit a dominant central peak surrounded by weaker extended emission. 
These compact continuum peaks likely mark the locations where disk-like structures may be present.

In this paper, we focus our analysis on the central source(s) in each region. 
A detailed identification and characterization of continuum sources 
across the full fields will be presented in a future paper.

\subsection{Extended Molecular Line Emission}
\label{sec:hotcore_lines}

Figure~\ref{fig:sample_lines} presents the moment 0 (integrated intensity) and moment 1 
(intensity-weighted velocity) maps of three molecular lines toward the ten sources. 
The lines shown are 
CH$_{3}$CN ($12_{4}-11_{4}$; $E_u/k = 183.1~\K$), 
SO$_{2}$ ($28_{3,25}-28_{2,26}$; $E_u/k = 403.0~\K$), 
and SiO ($5-4$; $E_u/k = 31.3~\K$), with detailed information listed in Table~\ref{tab:lines}.

The moment 0 maps show that CH$_{3}$CN and SO$_{2}$ emission is generally concentrated 
toward the central source, while also exhibiting significant extended components 
beyond the most compact continuum peaks. 
The SO$_{2}$ emission is typically more compact than that of CH$_{3}$CN,
as expected given its higher upper-state energy, 
which makes it more sensitive to warmer and more centrally concentrated gas.
In all sources, the moment 1 maps of CH$_{3}$CN and SO$_{2}$ display similar velocity gradients 
aligned in the same directions, 
indicating that the two species trace related kinematic structures 
on comparable spatial scales. 
These kinematic patterns appear to be dominated by large-scale rotation, 
consistent with 
QUARKS lower-resolution observations (QUARKS team, priv. comm.), 
although the presence of multiple systems introduce local distortions in several sources.

Notably, the emission peaks of CH$_{3}$CN and SO$_{2}$ in most sources do not coincide with the continuum peaks. 
In few cases
(e.g., I16484$-$4603, I17008$-$4040, and I17016$-$4124), 
signatures of self-absorption and/or absorption against the continuum are evident,
particularly for CH$_{3}$CN,
suggesting that these transitions primarily trace cooler, outer material, 
likely associated with infalling and rotating envelopes. 
As a result, these lines are not optimal tracers of the innermost disk regions 
(i.e., scales of a few hundred au).

The high-resolution SiO emission exhibits more diverse behavior. 
In some sources (e.g., I16484$-$4603 and I18134$-$1942), 
it appears compact and peaks at or very near the continuum position. 
In these cases, the SiO velocity gradients within the central few hundred au 
are broadly consistent with those traced by CH$_{3}$CN and SO$_{2}$. 
In contrast, in other sources (e.g., I17016$-$4124 and I18469$-$0132), 
the SiO emission is more extended or irregular, sometimes showing dips or absorption features toward the continuum peaks. 
In these cases, the velocity gradients are often oriented perpendicular to those seen in CH$_{3}$CN and SO$_{2}$, 
indicating significant contributions from outflows. 
In a few sources (e.g., I17008$-$4040 and I18517$+$0437), 
the SiO emission shows velocity gradients consistent with CH$_{3}$CN and SO$_{2}$ 
on the innermost scales, but transitions to outflow-dominated kinematics at larger radii.

\subsection{Water, Salt, and SiS Lines}
\label{sec:hotdisk_lines}

We search for emission from a vibrationally excited water line 
(H$_{2}$O $v_{2}=1$ $5_{5,0}-6_{4,3}$; $E_u/k = 3462$ K), 
as well as NaCl and SiS transitions in these sources
(see Table~\ref{tab:lines}). 
Because these lines are typically weak and can be contaminated by nearby transitions 
-- primarily from complex organic molecules (COMs) -- 
their direct identification from spectra alone is challenging.
However, these lines are expected to be spatially compact and to extend to higher velocities than typical COM emission. 
We therefore adopt a spatial--kinematic criterion for detection: a line is considered detected only if compact blueshifted and redshifted emission components, with spatial extents smaller than five beam sizes (typically $\lesssim 300$ au), are identified in close proximity to the continuum peak and located on opposite sides of it.
These components are defined by integrating over appropriate blueshifted and redshifted velocity ranges to 
produce moment 0 maps, and requiring at least five consecutive pixels above the $3\sigma$ noise level. 
Sources that exhibit emission on only one side, or show extended emission not clearly associated with the continuum peak, 
are classified as non-detections.

\begin{figure*}[ht!]
\centering
\includegraphics[width=\textwidth]{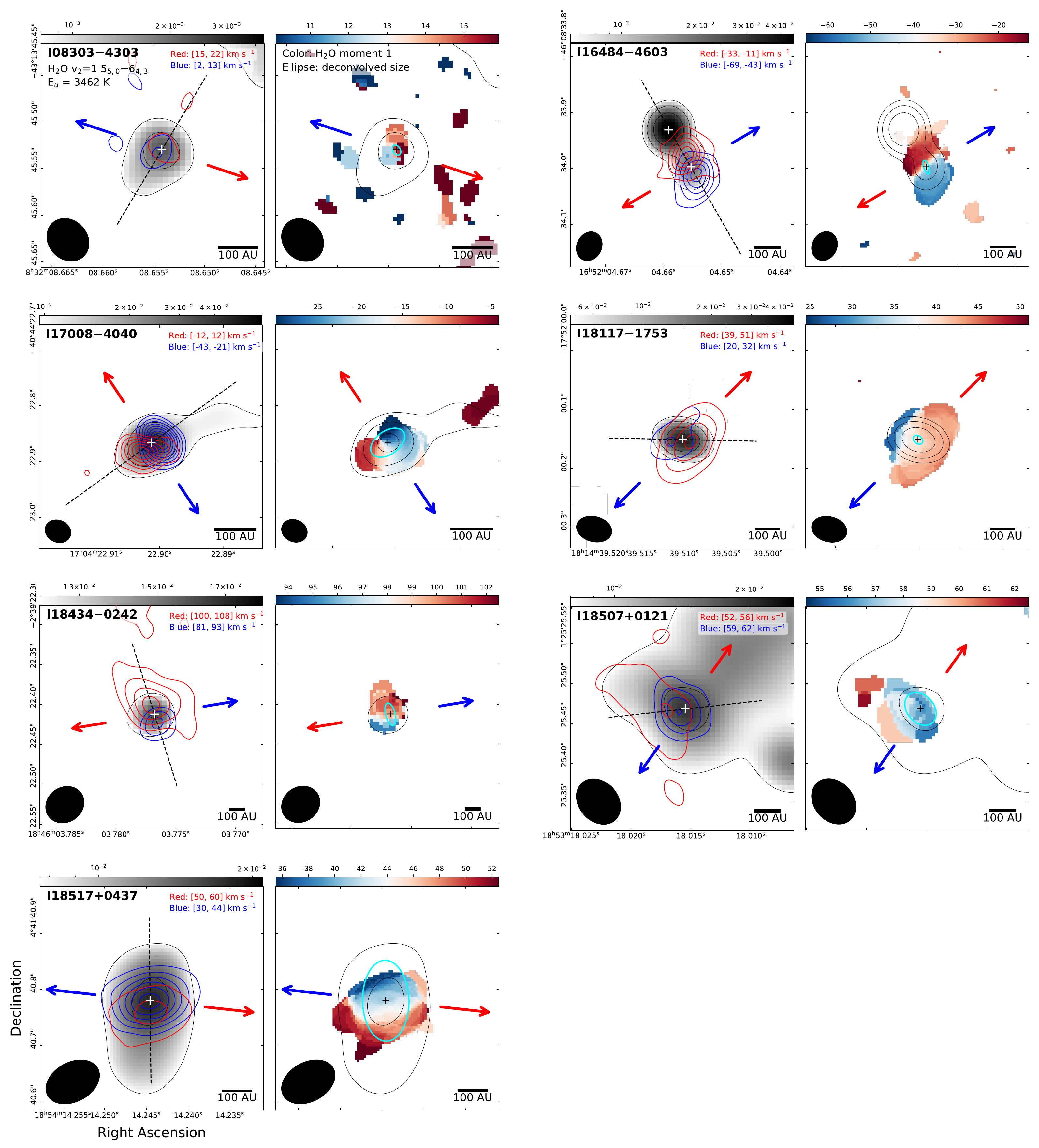}
\caption{Integrated blueshifted and redshifted emission maps and moment 1 maps of H$_2$O 
overlaid on the 1.3 mm continuum emission (grayscale and black contours). 
Red and blue arrows indicate the outflow directions. 
The integrated $v_{\rm lsr}$ ranges are labeled in each moment 0 panel.
Blue and red contours are plotted at levels of $3\sigma_{\rm area} \times n$ ($n = 1, 2, 3, \ldots$), where $\sigma_{\rm area}$ is the moment 0 map noise calculated following  $\sigma_{\rm area} = \sigma_\mathrm{chan} \sqrt{\Delta V_{\rm int}\,\delta V_{\rm chan}}$, where $\sigma_\mathrm{chan}$ is the rms noise per channel, $\Delta V_{\rm int}$ the integrated velocity range, and $\delta V_{\rm chan}$ is the channel width in velocity.
Black contours show the 1.3 mm continuum emission at levels of $10\sigma_{\rm conti} \times 2^{n}$ ($n = 5, 6, 7, \ldots$), 
except for I08303$-$4303, where $n = 2, 3$. 
$\sigma_{\rm conti}$ and $\sigma_{\rm chan}$ are listed in Table~\ref{tab:data}.
The cyan ellipses in the moment 1 panels mark the beam-deconvolved sizes of the central continuum emission.
The black dashed lines mark the PV cut paths. The color bars show the continuum intensity scales in Jy beam$^{-1}$ and the moment-1 velocity ranges in km s$^{-1}$.}
\label{fig:h2o}
\end{figure*}

We detect vibrationally excited H$_{2}$O emission within the central few hundred au in seven of the ten sources. 
Figure~\ref{fig:h2o} presents the integrated blue- and redshifted intensity maps, 
along with the intensity-weighted velocity (moment 1) maps, for these sources. 
In contrast to CH$_{3}$CN and SO$_{2}$, the water emission is highly compact and coincident with the continuum peaks, 
confined to scales of $\sim$100 au. In all cases, the emission exhibits clear velocity gradients, 
with blue- and redshifted components arranged symmetrically about the continuum peak.

The velocity gradients traced by the water line are broadly consistent with 
those of CH$_{3}$CN and SO$_{2}$ on inner scales in most sources, 
and are nearly perpendicular to the outflow directions identified from CO observations 
in the QUARKS project (QUARKS team, priv. comm.). 
The beam-deconvolved continuum morphologies 
(derived using the CASA {\it imfit} task) 
are shown as cyan ellipses in Figure~\ref{fig:h2o}. 
The directions of the H$_2$O velocity gradients are well aligned with the major axes of
the continuum structures, with the exceptions of I08303$-$4303 and I18507+0121. 
Note that, for I08303$-$4303, the deconvolved continuum size is too small 
relative to the synthesized beam 
to allow a reliable determination of the intrinsic position angle.
We therefore interpret the H$_2$O velocity gradients as signatures of rotation in dense structures on $\sim$100 au scales, 
most likely tracing disks around the massive protostars.

\begin{figure*}[ht!]
\centering
\includegraphics[width=170mm]{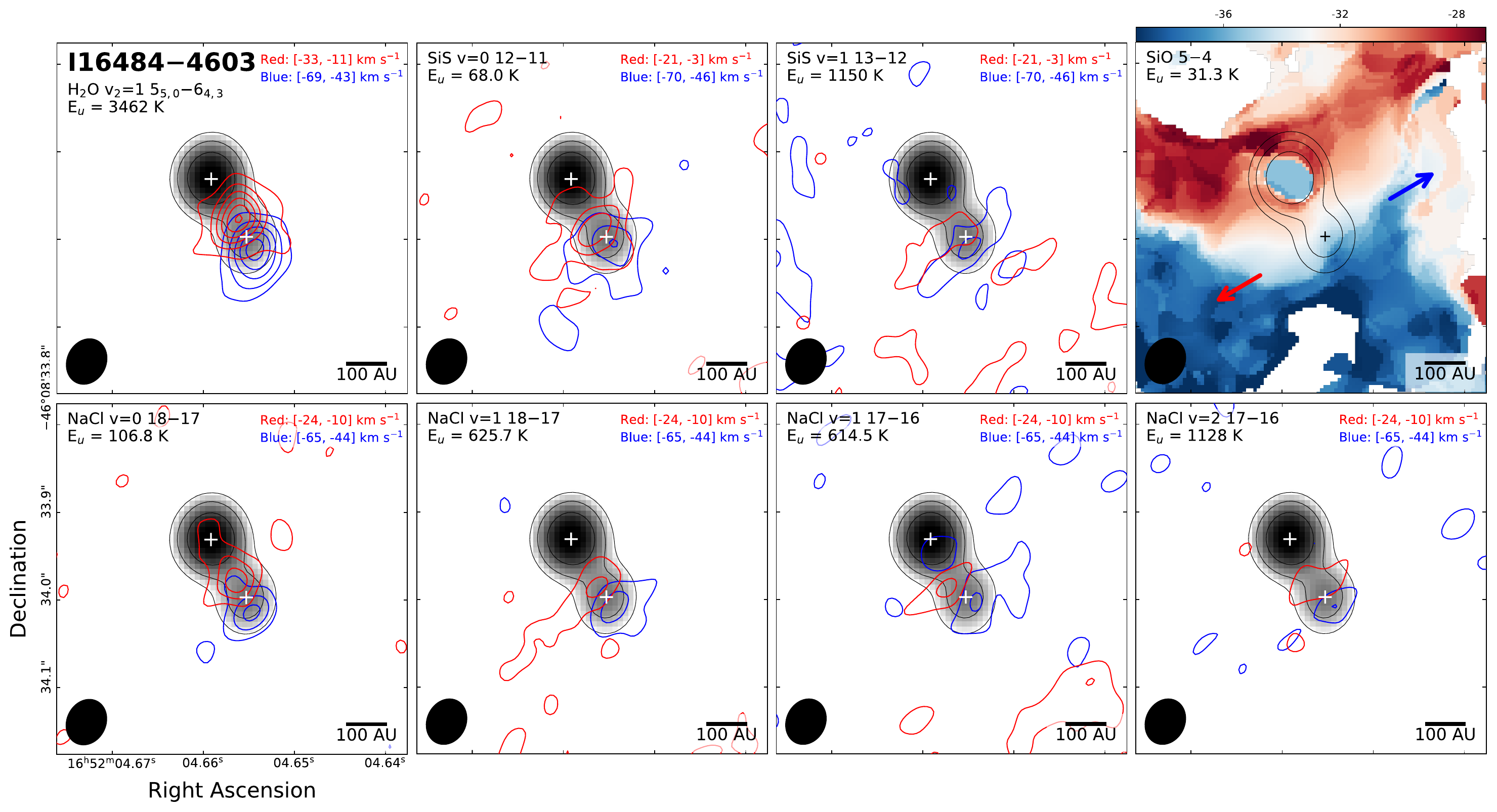}
\caption{Integrated blueshifted and redshifted emission maps of selected hot-disk molecular lines, together with the SiO moment-1 map, toward I16484$-$4603. The maps are overlaid on the 1.3 mm continuum emission, shown in grayscale and black contours.
The molecule names, transitions, upper-state energies, and integrated $v_{\rm lsr}$ ranges are labeled in each panel. 
Crosses mark the continuum peaks.
Blue and red contours are plotted at levels of $3\sigma_{\rm area} \times n$ ($n = 1, 2, 3, \ldots$), where $\sigma_{\rm area} = \sigma_{\rm chan} \sqrt{\Delta V_{\rm int}\,\delta V_{\rm chan}}$ and $\sigma_{\rm chan}$ is listed in Table~\ref{tab:data}.
The color bar shows the velocity range of the SiO moment-1 maps (km s$^{-1}$).}
\label{fig:16484}
\end{figure*}

\begin{figure*}[ht!]
\centering
\includegraphics[width=170mm]{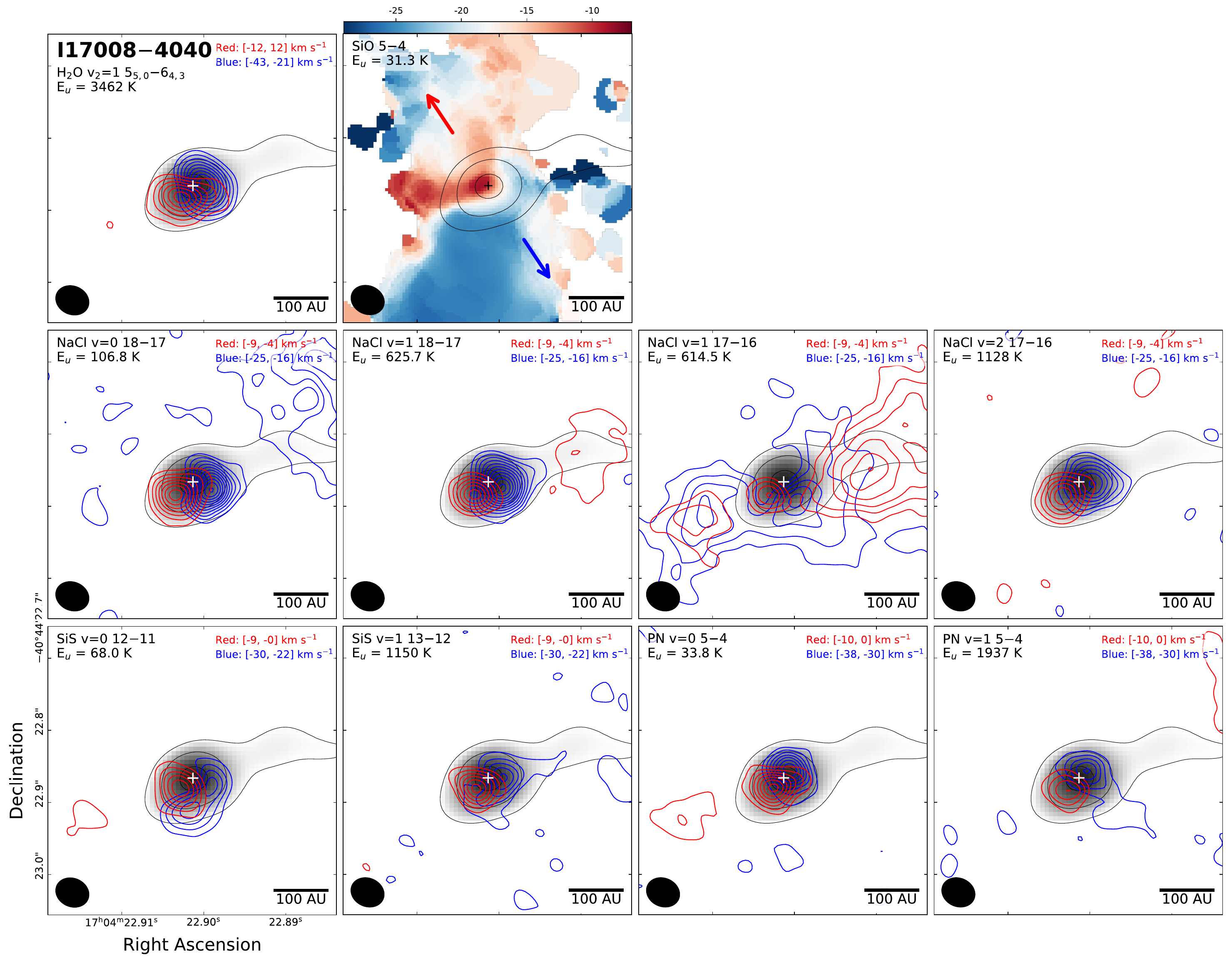}
\caption{Same as Figure~\ref{fig:16484}, but for I17008$-$4040.}
\label{fig:17008}
\end{figure*}

\begin{figure*}[ht!]
\centering
\includegraphics[width=170mm]{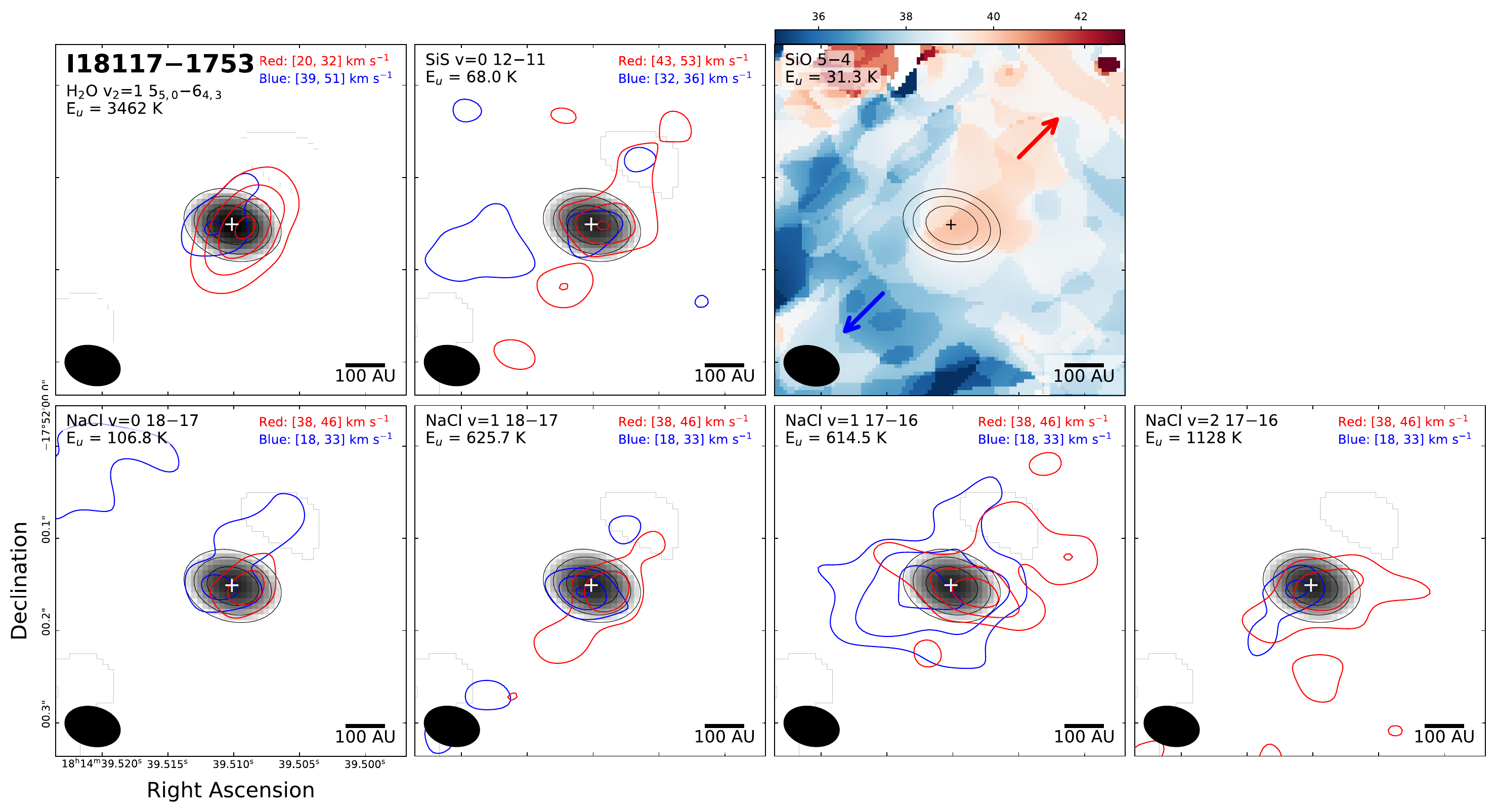}
\caption{Same as Figure~\ref{fig:16484}, but for I18117$-$1753.}
\label{fig:18117}
\end{figure*}

\begin{figure*}[ht!]
\centering
\includegraphics[width=170mm]{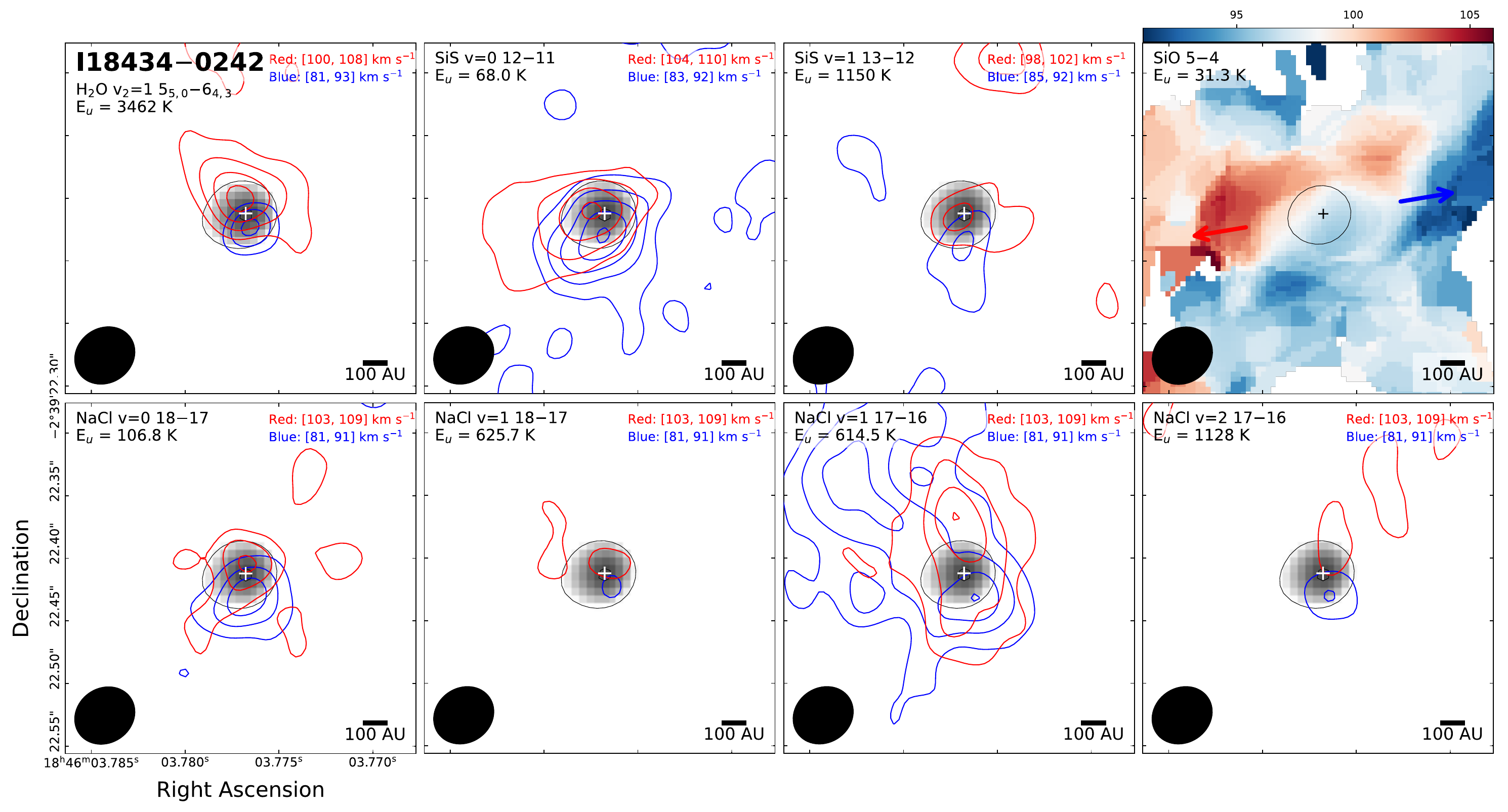}
\caption{Same as Figure~\ref{fig:16484}, but for I18434$-$0242.}
\label{fig:18434}
\end{figure*}

\begin{figure*}[ht!]
\centering
\includegraphics[width=130mm]{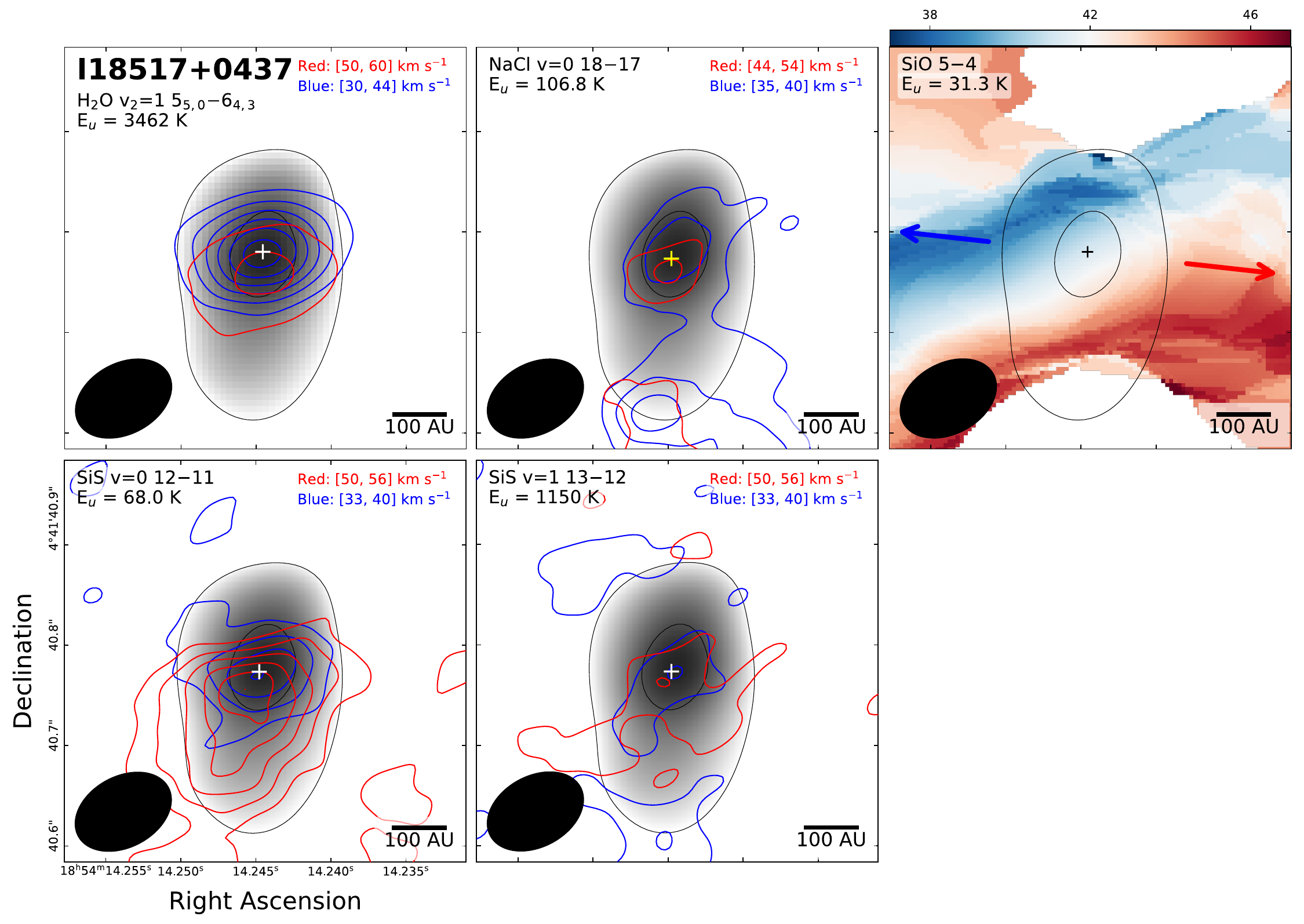}
\caption{Same as Figure~\ref{fig:16484}, but for I18517+0437.}
\label{fig:18517}
\end{figure*}

In addition, multiple NaCl and SiS transitions (listed in Table~\ref{tab:lines}) 
are detected toward five of the seven H$_{2}$O-detected sources. 
Figures~\ref{fig:16484}–\ref{fig:18517} show the integrated blue- and redshifted intensity maps of these transitions, along with the SiO 5$-$4 moment-1 maps. 
The detected NaCl and SiS lines include both pure rotational and vibrationally excited transitions, 
with $E_u/k$ ranging from 70 to 1100 K. 
These emissions are highly compact, coincident with the continuum peak, and spatially consistent with the H$_{2}$O emission.
They exhibit velocity gradients closely matching those seen in the water line. 
Some transitions appear to be contaminated by nearby COM lines, 
resulting in more extended emission; 
however, the central components remain consistent with the H$_{2}$O emission 
in terms of location, size, and kinematics. 
The detections of these disk-tracing lines are summarized in Table~\ref{tab:info}, 
and their peak intensities are listed in Table~\ref{tab:peak}. 
We discuss individual sources below.

\medskip

\noindent\textbf{\textit{I08303$-$4303.}}
The central region is resolved into a binary (or possibly higher-order multiple) system. 
Vibrationally excited H$_2$O emission is marginally detected at the $\sim 3\sigma$ level within a $\sim$100 au region 
around the northern component. 
The blue- and redshifted H$_2$O components lie on opposite sides of the continuum peak and 
exhibit a velocity gradient perpendicular to the outflow axis. 
Interestingly, this source corresponds to the second-brightest continuum peak and 
lacks hot core tracers such as CH$_{3}$CN and SO$_{2}$, 
which are instead concentrated toward a brighter southern source at a projected separation of $\sim$600 au.
In contrast, SiO emission is also concentrated toward the northern component; 
however, its velocity gradient on the innermost scales does not match that of H$_2$O and 
is likely dominated by outflow motions.
No H$_{2}$O emission is detected toward the southern source, and no NaCl or SiS emission is identified in both sources.

\medskip

\noindent\textbf{\textit{I16484$-$4603.}}
This source is also known as G339.88$-$1.26. 
Previous observations at lower angular resolution (comparable to our TM2 data) 
have revealed an ordered envelope-to-disk transition, with CH$_3$OH and H$_2$CO tracing the envelope, 
SO$_2$ and H$_2$S probing the inner envelope or envelope–disk transition zone, 
and SiO tracing the disk \citep{Zhang2019massive}. 
However, these observations constrained the disk size only to $\lesssim 300$ au.

Our high-resolution data resolve the central source into a binary system with a projected separation of $\sim$170 au. 
Molecular line emission is detected only toward the southern component, which has weaker continuum emission. 
Strong H30$\alpha$ emission (see Appendix \ref{sec:app_h30a}) 
is observed toward the northern component, 
indicating the presence of a highly compact ionized region that likely contributes significantly to its continuum emission. 
Much weaker H30$\alpha$ emission is also detected toward the southern component.
The H$_2$O line reveals a $\sim$200 au disk-like structure with a southwest–northeast velocity gradient, 
aligned with the binary axis and perpendicular to the CO outflow and SiO jet \citep{Zhang2019massive}. 
NaCl and SiS emission is detected at the $\sim 6\sigma$ level and exhibits similar kinematics.
All molecular lines -- including both ``hot-disk'' tracers and hot core tracers --
show velocity structures symmetric with respect to the southern source.

The SiO emission exhibits a velocity gradient consistent with that of H$_2$O and NaCl on disk scales, 
but also extends eastward along the outflow direction from the southern source. 
Notably, it continues to show velocity gradients perpendicular to the outflow axis, 
suggesting the presence of a rotating SiO wind, similar to that observed in Orion Source I \citep{2017NatAs...1E.146H}.

\medskip

\noindent\textbf{\textit{I17008$-$4040.}}
A well-resolved disk is traced by H$_{2}$O, NaCl, and SiS emission, 
showing the highest signal-to-noise ratio in the sample. 
PN ($5-4$ $v=0,1$) emission tracing a similar structure is also detected, 
making this the only source in the sample with such emission. 
All of these lines show a northwest–southeast velocity gradient.
The SiO emission exhibits a similar velocity gradient in the innermost region, 
but transitions to outflow-dominated kinematics at larger distances, 
with blueshifted emission extending toward the south.

\medskip

\noindent\textbf{\textit{I18117$-$1753.}}
This source is also known as W33A. 
\citet{Ginsburg2023} reported tentative detections of H$_{2}$O and NaCl using line stacking, 
without resolving the inner structure. 
Our observations resolve a disk-like structure with a clear east–west velocity gradient in 
H$_{2}$O, NaCl, and SiS emission. 
The improved detection likely results from the higher sensitivity of our data (2.4 K versus 6.5 K). 
While PN emission is present in the spectrum \citep{Ginsburg2023}, 
we do not identify a disk-like structure in this line.
The SiO emission is widespread across the central region but shows absorption toward the continuum peak. 
This absorption may arise from colder, low-velocity, and spatially extended SiO gas, 
possibly associated with cloud-cloud collision or filament convergence 
(F. Xu et al., submitted).
No clear velocity gradient is detected on disk scales.

\medskip

\noindent\textbf{\textit{I18434$-$0242.}}
This source is also known as G29.96. 
Previous observations did not reveal clear H$_{2}$O or NaCl emission or evidence of rotation \citep{Ginsburg2023}. 
Our data detect redshifted H$_{2}$O, NaCl, and SiS emission to the north of the continuum peak and blueshifted emission to the south, 
including both ground-state and vibrationally excited transitions.
The SiO emission is consistent with this velocity gradient on the innermost scales.
This inner-scale pattern is opposite to the red-/blue-shifted morphology of the large-scale SiO outflow, 
which shows red-shifted emission to the south \citep{2017A&A...602A..59C}.

\medskip

\begin{figure*}[ht!]
\centering
\includegraphics[width=16cm]{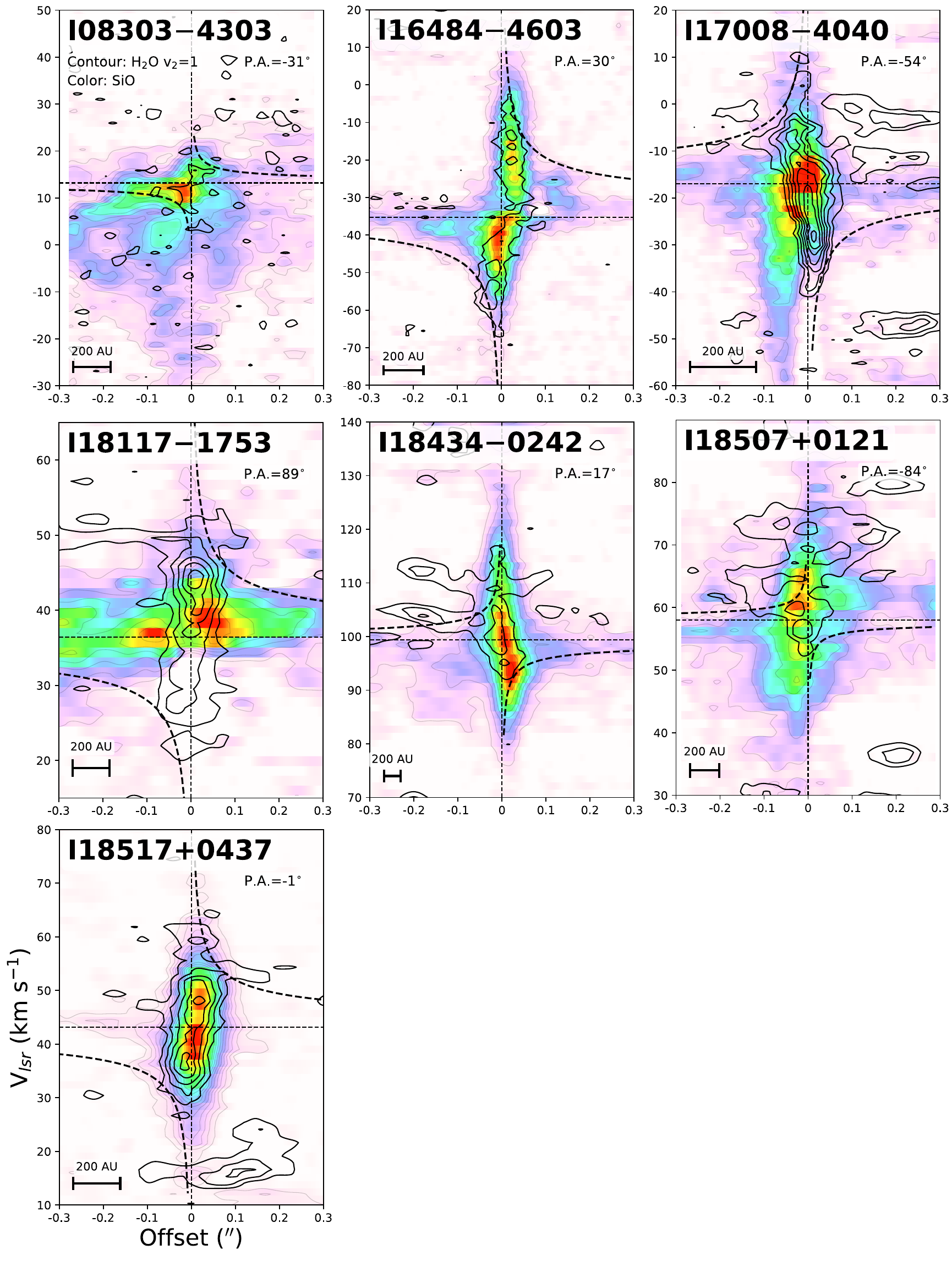}
\caption{Position–velocity (PV) diagrams of H$_{2}$O emission (black contours) 
overlaid on SiO emission (color scale) toward the hot-disk sources. 
The PV cuts are taken along the H$_{2}$O velocity-gradient direction of each source (see dashed lines in Figure \ref{fig:h2o}), with the position angles of the cuts presented in the upper-right corners. 
Dashed horizontal and vertical lines mark the systemic velocity 
and the central position, respectively.
Dashed curves show the Keplerian rotation curves 
corresponding to different central masses assuming an edge-on geometry.
For I16484$-$4603 and I17008$-$4040, the Keplerian curves are derived from double power-law fits to rotation curves measured along the PV-diagram edges.}
\label{fig:pv}
\end{figure*}

\noindent\textbf{\textit{I18507+0121.}}
This source is also known as G34.43.
The blue- and redshifted emission peaks of the H$_2$O line show only a small spatial separation, 
with a tentative southwest–northeast velocity gradient. 
The redshifted emission is partially contaminated by nearby spectral lines at $v_{\rm lsr}-v_{\rm sys} \approx 7$ and 15 km s$^{-1}$ 
(see Figure~\ref{fig:pv}). 
No NaCl or SiS emission is detected above the $3\sigma$ level. 
This source was previously classified as lacking a clear disk \citep{Ginsburg2023}; 
our detection likely benefits from improved resolution and lower noise (4.4 K).

\medskip

\noindent\textbf{\textit{I18517+0437.}}
H$_{2}$O, NaCl, and SiS emission all trace a disk-like structure on the innermost scale, 
although fewer transitions are detected. 
Among them, the vibrationally excited H$_{2}$O line shows the clearest north–south velocity gradient.

\section{Discussions}
\label{sec:dis}

\subsection{Water, Salt, and Silicon Compounds as Disk Tracers}
\label{sec:disk}

\subsubsection{Kinematic Patterns}
\label{sec:pvdiagram}

Figure~\ref{fig:pv} presents the position-velocity (PV) diagrams of the H$_2$O and SiO emission for the seven sources with detected hot-water lines. 
The PV cuts were extracted along the H$_2$O velocity gradients passing through the central continuum sources (dashed lines in Figure~\ref{fig:h2o}) 
with a width of $0.1^{\prime\prime}$.
The vibrationally excited H$_2$O emission is spatially compact, 
typically confined within a radius of $\sim0.05\arcsec$, 
but spans a wide velocity range.
In I16484$-$4603 and I17008$-$4040, the full velocity extent 
reaches $\sim 50-70~\kms$, 
while in the remaining sources it is typically $\sim20-30~\kms$
(see also the spectra in Appendix~\ref{sec:app_spec}).

In several sources (I17008$-$4040, I18117$-$1753, I18434$-$0242, and I18507+0121), 
the H$_2$O PV diagrams show contamination from unidentified spectral lines.
These contamination features can be recognized 
at velocity offsets of $v_{\rm lsr}-v_{\rm sys} \approx +7$, $+15$ and $-30~\kms$,
likely due to nearby complex organic molecule lines.
They can be distinguished from the H$_{2}$O emission by 
their extended spatial distribution (or offset from the center) and 
narrower velocity extent.
The same contamination lines were also seen in I16547$-$4247 (\citealt[]{Tanaka2020}).
A comprehensive line identification across the full spectral setup 
will be required in future work.

Clear velocity gradients are observed in the H$_2$O PV diagrams for most sources, 
except for I08303$-$4303, where the detection is marginal and 
the signal-to-noise ratio is insufficient to reveal a well-defined PV structure. 
Nevertheless, the moment maps still suggest the presence of compact, 
oppositely shifted components, 
and we therefore retain this source as a detection. 
In several sources (e.g., I16484$-$4603, I17008$-$4040, I18434$-$0242, and I18517+0437), 
the SiO kinematics broadly match those of H$_2$O. 
In I17008$-$4040, the blueshifted SiO emission shows two components: 
one consistent with the H$_2$O emission and 
another extending to higher velocities on the opposite side, the third quadrant in the PV diagram,
likely associated with the outflow. 
In I18117$-$1753, the absence of blueshifted SiO emission 
within the central $0.1\arcsec$ is plausibly due to self-absorption. 
When the signal-to-noise ratio is sufficiently high, 
both H$_{2}$O and SiO exhibit similar velocity patterns consistent with rotation.

\begin{figure*}[ht!]
\centering
\includegraphics[width=18cm]{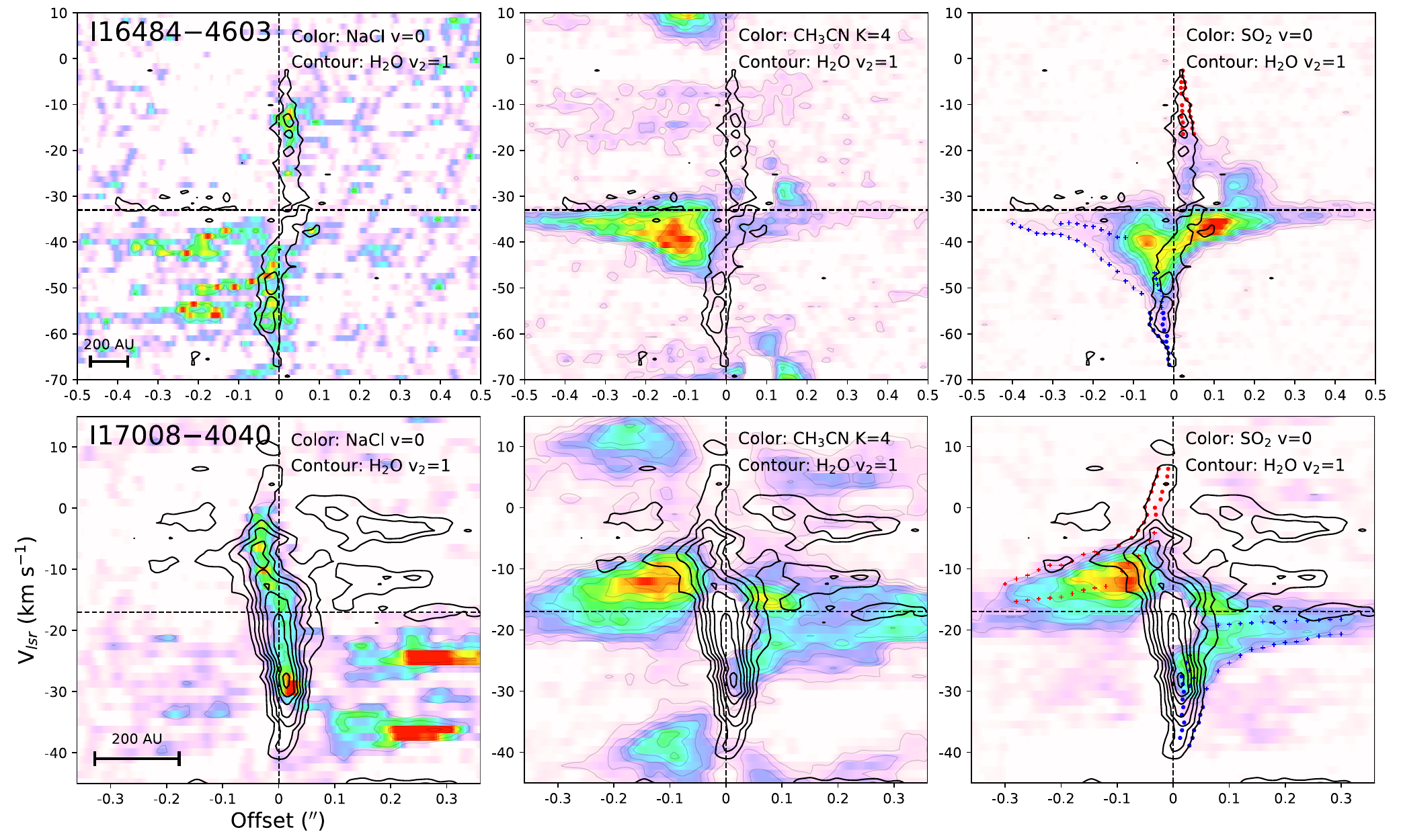}
\caption{Position–velocity (PV) diagrams of H$_{2}$O emission 
(black contours) overlaid on NaCl (18$-$17) (left), CH$_{3}$CN ($12_{4}-11_{4}$) (middle), 
and SO$_{2}$ ($28_{3,25}-28_{2,26}$) (right) emission (color scale)
toward I16484$-$4603 (upper panels) and I17008$-$4040 (lower panels). 
The PV cuts are taken along the velocity-gradient direction of each source
(see dashed lines in Figure \ref{fig:h2o}).
The blue and red circles mark the velocity-position measurements 
derived from the H$_2$O PV diagrams for the blueshifted and redshifted emission, 
respectively (see text for details). 
The crosses indicate analogous measurements from the SO$_2$ line. 
Measurements are made for both the ridge and edge of the PV diagrams 
and are used in Figure~\ref{fig:pv_curves}.}
\label{fig:pv_lines}
\end{figure*}

To further assess the kinematic origin of the hot disk lines, 
we compare the PV structures of H$_2$O $v_2=1$ and NaCl lines 
with hot core molecular tracers such as CH$_3$CN and SO$_2$, 
for the two sources with high signal-to-noise ratios and 
well-separated red- and blueshifted components, 
I16484$-$4603 and I17008$-$4040 (Figure~\ref{fig:pv_lines}). 
In both sources, the NaCl emission is compact and centrally concentrated, 
closely resembling the H$_2$O emission. 
Their velocity structures are also broadly consistent, 
exhibiting the characteristic signatures of disk rotation.

In contrast, the CH$_3$CN and SO$_2$ emission differs markedly from H$_2$O and NaCl. 
Although these tracers often show organized velocity gradients, 
they are more spatially extended and do not reach the high velocities observed in H$_2$O.
Moreover, they exhibit weak or absent emission at the central position and
display more complex velocity structures. 
Their PV morphologies are therefore inconsistent with the compact, disk-like 
kinematics traced by H$_{2}$O and NaCl, 
suggesting that they primarily trace larger-scale structures 
such as the rotating envelope.

To quantitatively characterize the rotational kinematics traced by 
the hot-disk and hot-core lines, we extract rotation curves 
(rotation velocity versus radius) from the PV diagrams of the 
H$_2$O and SO$_2$ lines toward I16484$-$4603 and I17008$-$4040,
following the methods widely used in low-mass protostellar disk studies
\citep[e.g.,][]{2015ApJ...812...27A,2020ApJ...893...51S,2023ApJ...951...10V}. 
Rotation curves are commonly derived either from the emission boundary 
(the ``edge'' method) or from the locus of peak emission 
(the ``ridge'' method) in PV diagrams. 
We employ both approaches in our analysis.

\begin{figure*}[ht!]
\centering
\includegraphics[width=18cm]{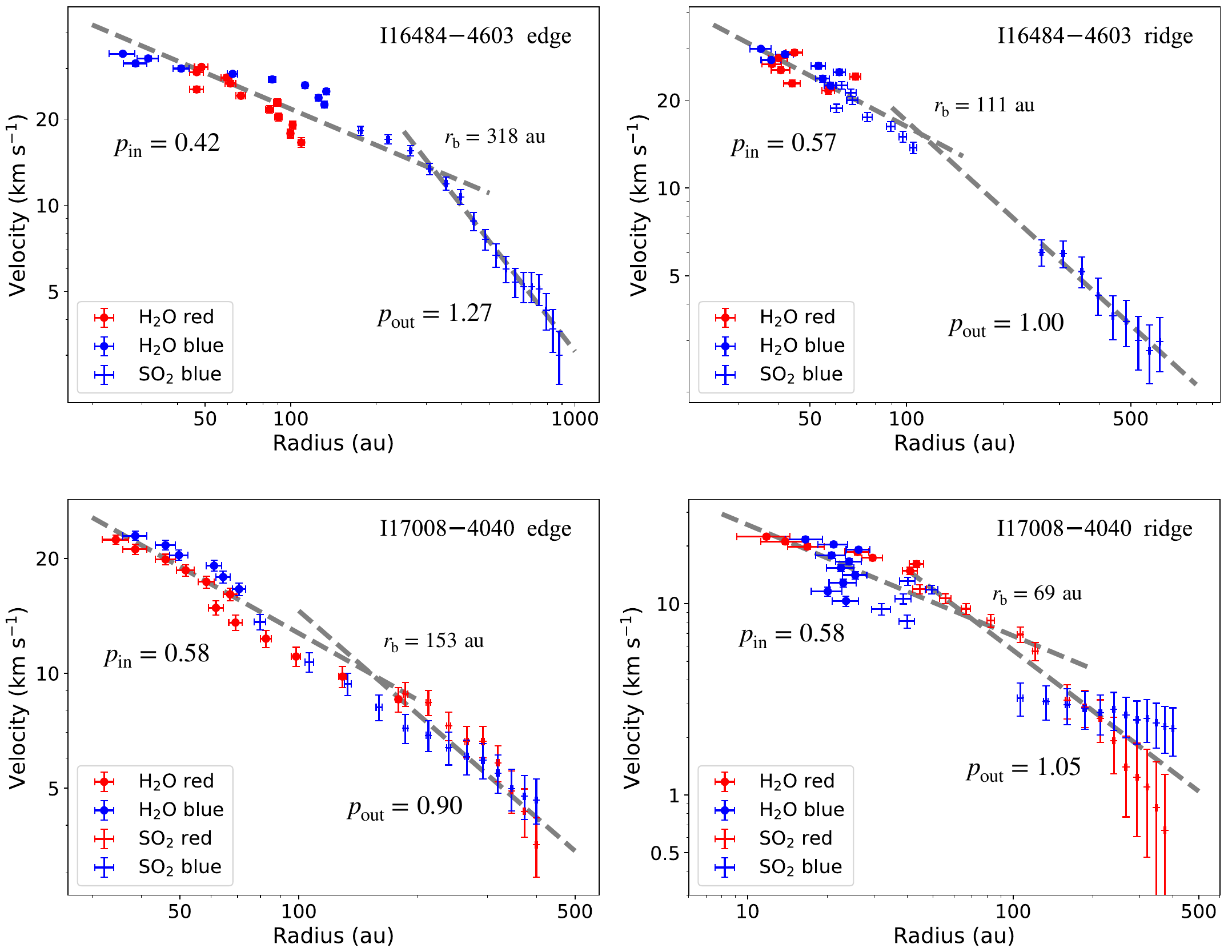}
\caption{
Rotation curves derived from the PV diagrams 
along the disk major axes of I16484$-$4603 and I17008$-$4040 
(see Figure~\ref{fig:pv_lines}). 
Circles and crosses denote measurements from the H$_2$O ($v_2=1$; $5_{5,0}-6_{4,3}$) 
and SO$_2$ ($28_{3,25}-28_{2,26}$) lines, respectively. 
Blue and red symbols represent the blueshifted and redshifted emission. 
Measurements derived from both the ridge and edge of the PV diagrams are shown. 
The dashed lines indicate the best-fit double power-law models, 
with fitted parameters listed in Table~\ref{tab:pv_curve}. 
The redshifted SO$_2$ measurements are excluded from the fit 
because they are contaminated by emission near the companion source. 
Velocities are given relative to the systemic velocity.}
\label{fig:pv_curves}
\end{figure*}

For the high-velocity H$_2$O and SO$_2$ emission, 
we determine the edge points in each velocity channel 
as the largest position offsets with emission above the $3\sigma$ level. 
The ridge points are computed as the intensity-weighted mean position 
at each velocity.
For the low-velocity SO$_2$ emission, 
the edge points are instead defined as the largest velocity offsets 
with emission above the $3\sigma$ level at each position offset. 
The positions are sampled every five pixels, 
corresponding to approximately half a synthesized beam. 
The ridge points are calculated as the intensity-weighted mean velocity
at each position.
The extracted edge and ridge points are overlaid on the 
H$_2$O and SO$_2$ PV diagrams in Figure~\ref{fig:pv_lines}, 
while the corresponding rotation curves are presented in Figure~\ref{fig:pv_curves}.

For both sources and for both extraction methods, 
the rotation curves exhibit a clear break in slope, 
with a relatively shallow profile in the inner region 
and a steeper profile at larger radii. 
To quantify this behavior, we fit the rotation curves with a double power-law model, 
allowing both the inner and outer power-law indices to vary freely. 
The break radius, $r_b$, is also treated as a free parameter. 
To account for small velocity offsets between the blueshifted and redshifted emission, 
we additionally allow the systemic velocity, $v_\mathrm{sys}$, 
to vary slightly during the fitting procedure. 
The best-fit models are shown in Figure~\ref{fig:pv_curves}, 
and the fitted parameters are listed in Table~\ref{tab:pv_curve}.

The fitted inner power-law indices are close to 0.5 in both sources, 
while the outer indices are approximately 1. 
For Keplerian rotation, one expects $v_\mathrm{rot}\propto r^{-0.5}$, 
whereas angular-momentum-conserving infall with rotation 
follows $v_\mathrm{rot}\propto r^{-1}$. 
The fitted slopes therefore suggest that both sources consist of 
an inner Keplerian disk embedded within an infalling, rotating envelope. 
To estimate the dynamical masses, we also perform fits with the inner 
power-law index fixed at 0.5, corresponding to a Keplerian rotation profile,
\begin{equation}
v_\mathrm{rot}=\sqrt{\frac{GM_\ast}{r}}\sin i.
\end{equation}
The resulting parameters are also listed in Table~\ref{tab:pv_curve}.

Using the edge method, we derive dynamical masses of 46 and 20~$M_\odot$ 
for I16484$-$4603 and I17008$-$4040, respectively, 
while the ridge method yields corresponding masses of 33 and 6.2~$M_\odot$. 
The transition radii derived from the edge method are 318 and 153~au, 
compared to 111 and 69~au from the ridge method 
for I16484$-$4603 and I17008$-$4040, respectively. 
The systematically lower masses and smaller transition radii obtained 
from the ridge method are consistent with previous studies 
showing that intensity-weighted ridge measurements tend to 
underestimate the intrinsic rotation velocity 
because of beam smearing and contamination from lower-velocity emission, 
whereas the edge method more closely traces the highest-velocity Keplerian rotation 
and generally provides a more reliable estimate of the central mass 
\citep[e.g.,][]{Seifried2016MNRAS,Aso2020ApJ}. 
The true dynamical masses and transition radii are therefore likely closer to the values derived from the edge method.

The transition radii can be interpreted as characteristic disk radii. 
The H$_2$O emission is largely confined within these radii, 
whereas the SO$_2$ emission is predominantly detected at larger radii, 
although it also extends into the outer disk. 
This result supports the interpretation that 
H$_2$O preferentially traces the innermost Keplerian disk, 
while SO$_2$ primarily traces the surrounding infalling-rotating envelope.
The importance of the hot disk tracers lies not primarily in 
obtaining a different estimate of the stellar mass, 
but in distinguishing the compact Keplerian disk from the surrounding 
infalling-rotating envelope. 
Without such tracers, the rotation traced by molecules 
such as SO$_2$ could be misinterpreted as a substantially larger Keplerian disk, 
leading to an overestimate of the disk radius.

\begin{figure*}[ht!]
\centering
\includegraphics[width=0.8\textwidth]{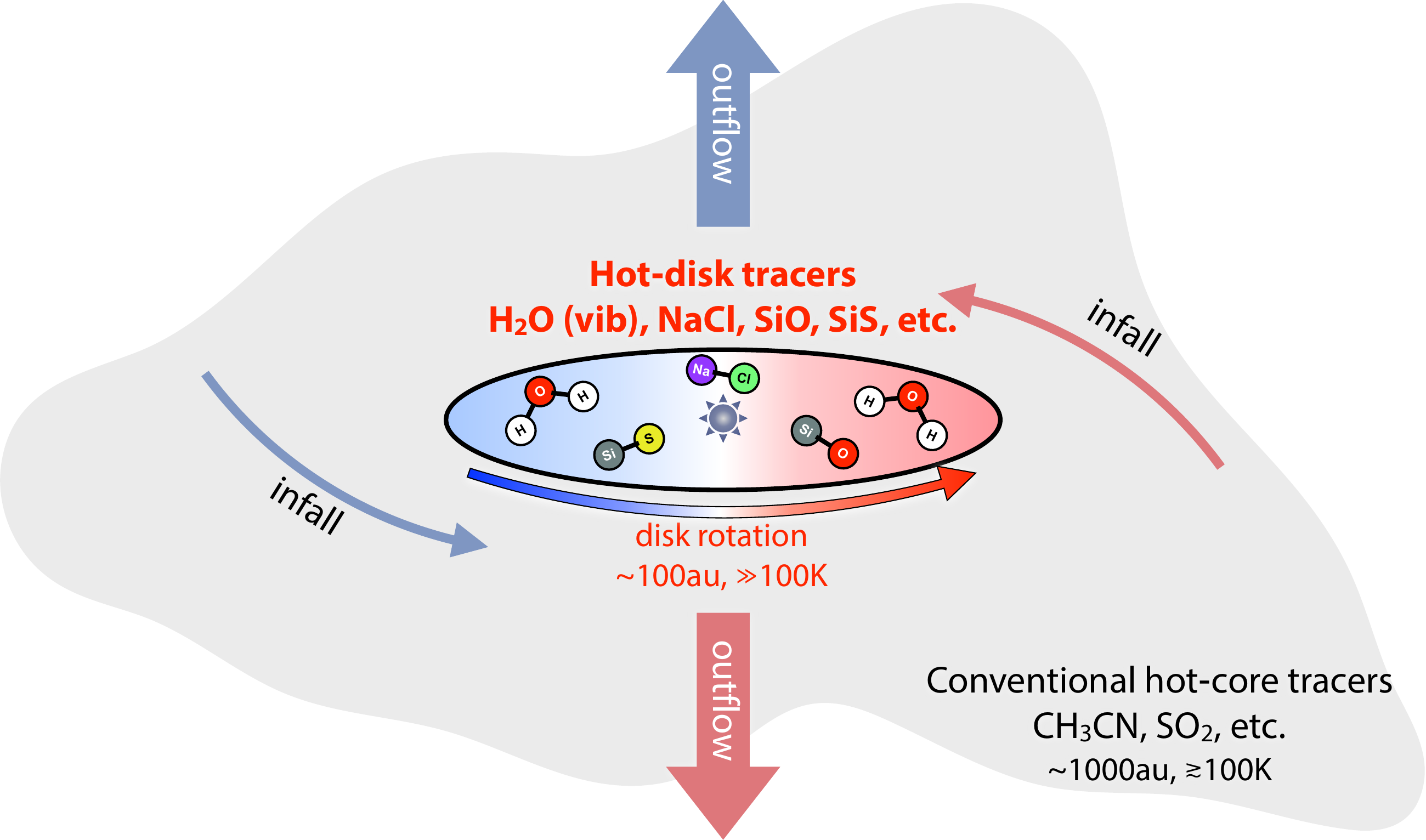}
\caption{Schematic illustration of the hot inner regions of massive protostellar systems,
highlighting the hierarchical structure and associated chemical tracers. 
Conventional hot-core tracers (e.g., CH$_3$CN and SO$_2$) primarily probe rotation on envelope scales 
($\sim$1000 au), whereas vibrationally excited H$_2$O and refractory species (e.g., NaCl, SiO, and SiS) 
trace compact, disk-scale rotation ($\sim$100 au) closer to the central protostar.}
\label{fig:schematic}
\end{figure*}

Taken together, these results support a hierarchical picture of 
rotational kinematics in massive protostellar systems. 
CH$_3$CN and SO$_2$ primarily trace rotation on envelope scales, 
whereas vibrationally excited H$_2$O and refractory species such as NaCl and SiS 
(and, in some cases, SiO) probe more compact, disk-scale rotation 
closer to the central protostar (see Figure \ref{fig:schematic}). 
Our results therefore indicate that vibrationally excited H$_2$O, NaCl, and SiS 
are powerful tracers of the innermost $\sim$100 au around massive protostars 
and provide a more direct probe of disk kinematics than traditional hot-core molecules. 
This interpretation is consistent with previous studies of hot-disk sources 
such as Orion Source I and I16547$-$4247 
\citep{2019ApJ...872...54G,Tanaka2020,Ginsburg2023}, 
and is further strengthened by the HOTDISK survey, 
which demonstrates that such disk-tracing signatures may be 
common among massive protostars.

\subsubsection{Origin of the ``Hot-disk'' Pattern}
\label{sec:origin}

Such spatial and kinematic patterns 
can be understood in terms of molecular origin and excitation. 
Molecules such as CH$_3$CN and SO$_2$ 
are commonly associated with warm envelope and/or envelope-disk interface gas, 
where species released from icy grain mantles remain abundant. 
In the case of SO$_2$, its abundance may also be enhanced by subsequent warm 
gas-phase chemistry following the release of sulfur-bearing precursors such as SO, 
rather than by direct desorption alone 
\citep[e.g.,][]{1998A&A...338..713H,2009ARA&A..47..427H,2021A&A...653A.159V}. 
By contrast, species such as NaCl, SiO, and SiS are preferentially associated 
with regions closer to the central protostar, 
where higher temperatures and energetic processing of dust are required, 
although the dominant release mechanism 
(e.g., dust destruction versus other high-temperature processing) remains uncertain. 
The compactness of the vibrationally excited H$_2$O emission 
may be driven not only by its high excitation energy ($E_u = 3462$ K), 
but also that radiative pumping may play an important role and 
non-LTE effects are likely significant. 
As a result, this H$_2$O $v_2=1$ transition may be highly sensitive to strong infrared radiation fields 
and preferentially traces the hottest inner regions. 
Therefore, the observed spatial distribution of this line may 
primarily reflect excitation conditions rather than the intrinsic distribution of water vapor.

Previous studies of Orion Source I suggest that NaCl primarily traces the disk rotation, 
vibrationally excited H$_{2}$O traces the disk and possibly the base of a disk wind, 
and SiO and SiS are more closely associated with the wind-launching region 
\citep{2017NatAs...1E.146H,2024ApJ...974..150W}. 
In this context, the observed velocity gradients may partly arise from a disk wind 
that remains co-rotating with the disk. 
However, given the current angular resolution, 
we cannot resolve the vertical structure of these components. 
In this work, we therefore do not distinguish between disk midplane, 
disk surface, and wind base. 
Instead, we interpret these tracers as probing compact, rotating gas 
closely associated with disk accretion.

Besides the two sources with rotation curve fitting,
we estimate the central masses and systemic velocities 
by visually matching Keplerian rotation curves to the edges of the H$_2$O PV diagrams
for other sources assuming edge-on Keplerian rotation. 
The model curves follow $v_{\rm rot}(r) = \sqrt{GM_\star/r}$ for an inclination angle of 
$i=90^{\circ}$, where $r$ is taken as the projected offset from the continuum peak along 
the PV cut. The curves are normalized by matching the observed velocity at the outer edge 
of the emission in the PV diagram, and the corresponding radius is used to determine the 
enclosed mass.
The dynamical mass estimation here contains large uncertainties.
First, the masses derived under the edge-on assumption can only provide lower limits. 
Second, these estimates are based only on simple visual matching to the lowest PV contours
(emission edges) rather than detailed kinematic modeling,
and strongly affected by the signal-to-noise ratio (e.g. I08303$-$4303).
Furthermore, the inferred masses may also be affected by line contamination in 
some sources (e.g., I18434$-$0242 and I18507+0121).
In sources where SiO shows clear outflow contributions 
(e.g., I16484$-$4603), the observed velocity structure may be further 
contaminated by outflow-related motions in addition to rotation. 
Therefore, the derived dynamical masses should be treated as reference values
with substantial uncertainties.
Protostellar masses based on infrared spectral energy distribution (SED) fitting
are also provided in Table~\ref{tab:info} for reference 
(assuming single source dominating the infrared emission; see Appendix \ref{sec:app_sed}).

\subsubsection{Excitation Conditions}
\label{sec:excitation}

\begin{figure*}[ht!]
\centering
\includegraphics[width=0.49\textwidth]{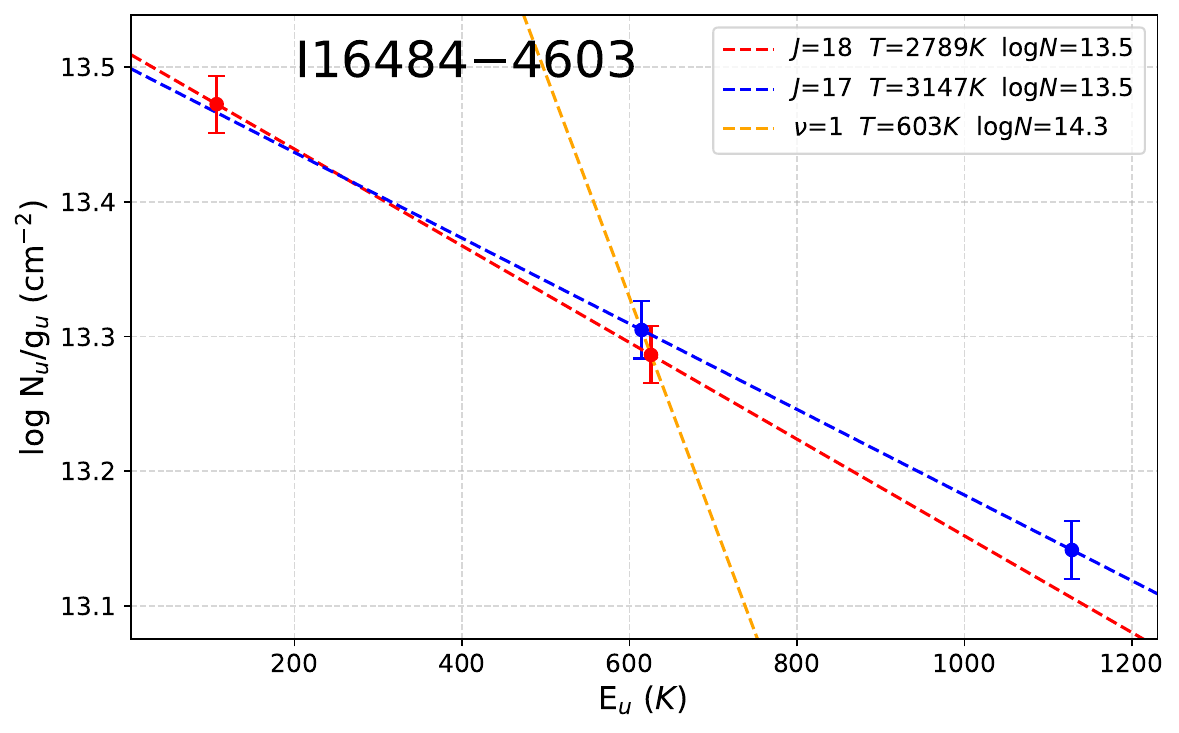}
\includegraphics[width=0.49\textwidth]{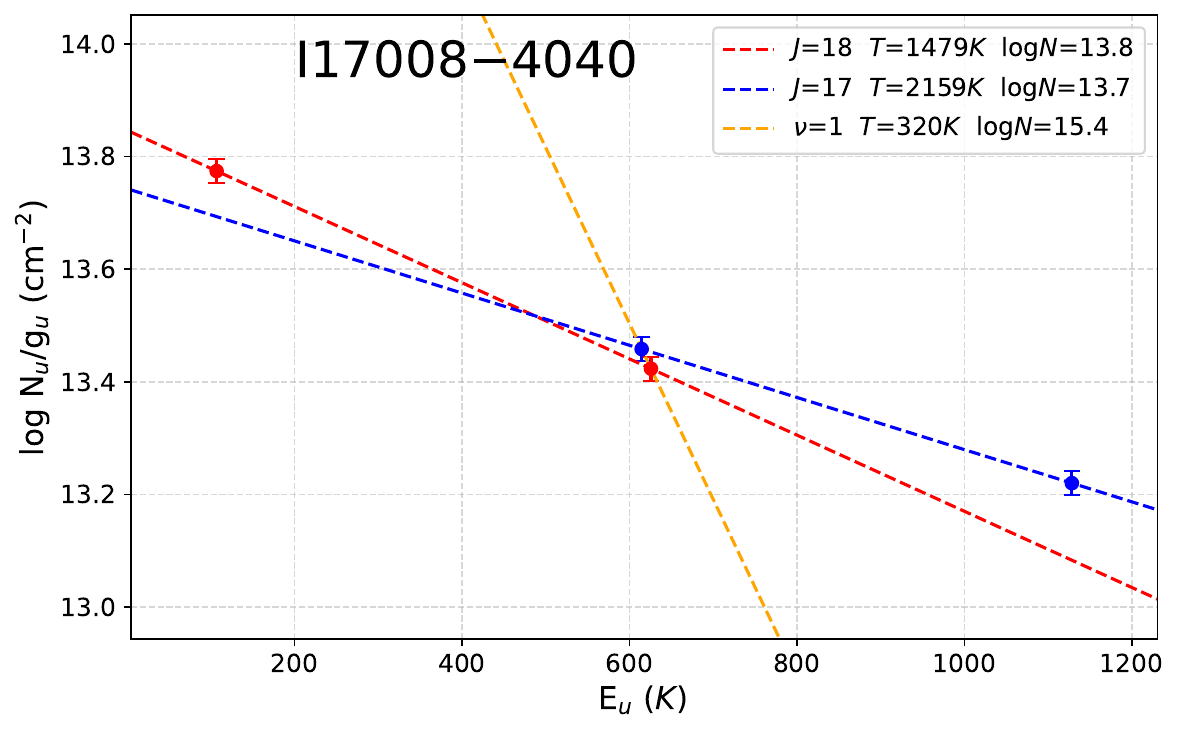}
\caption{Rotational and vibrational diagrams derived from NaCl lines toward I16484$-$4603 and I17008$-$4040.
Upper state column densities $N_u$ of each transition are measured within the central beam-sized region.
Rotational and vibrational temperatures calculated using different sets of lines
are shown in each panel.}
\label{fig:rot_vib}
\end{figure*}

Assuming local thermodynamic equilibrium (LTE), 
we construct rotational and vibrational diagrams for NaCl toward the two sources 
with the highest signal-to-noise ratios 
(I16484$-$4603 and I17008$-$4040; Figure~\ref{fig:rot_vib}). 

The derived rotational temperatures are $\sim$600~K and $\sim$300~K, respectively, 
whereas the vibrational temperatures inferred from the vibrationally excited transitions 
are higher, of order $\sim$3000~K and $\sim$1500--2000~K. 
These results are broadly consistent with previous studies 
\citep{2019ApJ...872...54G,Ginsburg2023}. 
The physical interpretation of the vibrational temperature depends on the excitation mechanism. 
If the excitation is dominated by collisions in dense and hot gas, the vibrational temperature 
may reflect the kinetic temperature. 
In contrast, if radiative pumping is important, the vibrational temperature can exceed the 
kinetic temperature and instead trace non-LTE excitation effects. 
In this case, the actual gas temperature is likely closer to the rotational temperature 
of a few hundred K. 
We note that, 
because the vibrational temperatures are estimated from transitions that are 
relatively close in upper-state energies, 
the derived rotational temperatures are sensitive to uncertainties in the 
line-flux ratios and should be regarded as indicative rather than precise
measurements.
However, the large differences between the rotational and vibrational temperatures remain significant.
We therefore interpret the high vibrational temperatures mainly as evidence for non-LTE 
excitation, suggesting that NaCl is not in LTE and that radiative excitation likely plays an important role.

Alternatively, the apparent excitation temperatures inferred from the rotational and vibrational diagrams may be biased by radiative transfer effects \citep{Goldsmith99}. 
In particular, when the optical depths of the molecular lines become non-negligible,
the contrast between transitions with different excitation energies is suppressed relative to that expected from their level populations,
which can flatten the slope of the diagrams and lead to an overestimation of the excitation temperature.
The presence of bright dust continuum emission may introduce additional uncertainty in the derivation of level populations from the observed line intensities.
Given the high continuum brightness temperatures of these sources 
(185 K and 596 K; Table~\ref{tab:peak}), 
the dust emission may approach optical depth unity. 
In addition, dust scattering 
-- suggested to be important even in massive protostellar disks \citep{Girart18,Yamamuro25} -- 
may further modify the emergent line intensities \citep{Bosman21}. 
A self-consistent treatment that simultaneously models dust continuum 
and molecular line radiative transfer 
will therefore be necessary to fully assess these effects.
Given the limited number of detected transitions, 
a more robust determination of the excitation conditions will 
require observations of additional lines.

\subsection{Detection Conditions of Hot-Disk Patterns}
\label{sec:detection}

In our sample of ten massive protostars, 
we detect vibrationally excited H$_2$O emission in seven sources. 
In these objects, the emission appears as compact red- and blueshifted components 
located on opposite sides of the central continuum peak, 
consistent with disk-like kinematics. 
In addition, five of these sources exhibit multiple NaCl and SiS transitions 
with similarly distributed red- and blueshifted emission, 
further supporting the presence of compact rotating structures. 
As discussed in the previous section, these sources most likely harbor 
accretion disks on scales of $\sim$100 au.

Among the seven hot-disk sources identified here, 
one (I18117$-$1753) was previously classified as a tentative detection by \citet{Ginsburg2023}, 
while two others (I18434$-$0242 and I18507+0121) were observed in that study without detection of 
either the relevant lines or disk-like kinematics. 
The remaining four sources are newly reported in this work. 
Our observations therefore significantly expand the sample of massive protostars 
with detected vibrationally excited H$_2$O, NaCl and SiS lines 
(i.e., hot-disk patterns),
reinforcing their effectiveness as disk tracers.

We emphasize that these hot-disk tracers are both spatially compact and 
intrinsically faint, requiring high angular resolution and sensitivity for detection. 
The identification of hot-disk patterns in sources previously reported as non-detections 
is likely attributable to the improved sensitivity of our observations 
(by a factor of $\gtrsim 2$), 
as the angular resolutions are comparable. 
For the three sources without disk detections, 
the emission may simply fall below our current sensitivity limits. 
In addition, insufficient angular resolution can lead to severe beam dilution and 
confusion with envelope emission.
In the detected sample, the typical size of the structures traced by 
vibrationally excited H$_2$O is $\lesssim 200$ au, which is generally comparable to that of the high-resolution continuum emission.
For the most distant source in our sample, I18469$-$0132 (4.9 kpc), 
the synthesized beam corresponds to $\sim$293 au, 
likely to be larger than the expected disk size. 
Such a beam can significantly dilute the emission and 
hinder the identification of disk-like kinematics. 
Based on our results toward disk-detected sources, 
an angular resolution corresponding to $\sim$100 au and 
a line sensitivity of $\lesssim 2.5$ K are required 
to robustly detect these disk-tracing lines.

We caution that our sample is small and biased, 
as the targets were preselected to exhibit clear CH$_3$CN rotational signatures 
and centrally concentrated SiO emission. 
Such sources are more likely to host disks, 
and the presence of refractory species may therefore be enhanced. 
Nonetheless, there may also be an intrinsic dependence on source properties. 
For example, among the seven disk-detected sources, 
only two exhibit H30$\alpha$ emission, 
whereas two of the three non-detected sources 
(including the northern component of I16484$-$4603) 
show strong H30$\alpha$ emission. 
Since H30$\alpha$ traces photoionized gas, 
this tentative anti-correlation suggests that hot molecular disks traced by 
``hot-disk'' molecules may preferentially exist 
at earlier evolutionary stages, before significant ionization disrupts or 
photoevaporates the inner disk. 
However, the current sample size is insufficient to establish this trend conclusively.
This is also consistent with \cite{Ginsburg2023}, who found no clear correlation or
anti-correlation between RRL emission and vibrationally excited H$_2$O emission, suggesting
that larger samples are required to determine whether the apparent anti-correlation is real.

In summary, for a sample selected to exhibit clear rotational signatures on core scales, 
our observations indicate that $\sim$70\% (or possibly higher) of sources 
show hot-disk chemical patterns consistent with rotating disks embedded 
within larger-scale rotating envelopes.
Given that larger-scale rotating cores are commonly seen around massive protostars,
these results suggest that such hot disks may not be rare, 
but instead may be common in massive star-forming regions, 
provided that sufficient sensitivity and angular resolution are achieved. 
This further supports a scenario in which rotationally supported disks 
are a natural outcome of massive star formation, 
at least in systems with well-developed rotating envelopes.


\section{Summary}
\label{sec:sum}

As part of the HOTDISK project, we present high-angular-resolution ($\sim0.05^{\prime\prime}$) 
ALMA Band~6 observations of ten massive protostellar sources, 
aimed at identifying the ``hot-disk'' chemical pattern in their innermost regions. 
Our main findings are summarized as follows:

\begin{enumerate}

\item Vibrationally excited H$_2$O ($5_{5,0}–6_{4,3}$, $v_{2}=1$; $E_u/k = 3462$ K) emission
is detected toward 7 out of 10 sources, corresponding to a detection rate of $\sim$70\%. 
In all detected cases, the emission is spatially compact ($\lesssim$100–200 au) 
and exhibits clear velocity gradients approximately perpendicular to the outflow directions,
strongly indicating rotating disk-like structures on $\sim$100 au scales.
Comparison with Keplerian rotation curves suggests 4 out of 7 systems have masses $\gtrsim 10~M{\odot}$, 
confirming that these systems host forming massive stars.

\item NaCl and SiS emission is detected toward 5 of the 7 H$_2$O-detected sources. 
These species show kinematics consistent with those of 
the vibrationally excited H$_2$O emission, 
reinforcing their association with compact, rotating inner structures. 

\item In contrast, commonly used hot-core tracers such as CH$_3$CN and SO$_2$ 
primarily trace larger-scale rotating envelopes rather than the inner disks. 
Although SiO emission can be centrally concentrated and, in some cases, 
exhibits kinematics consistent with disk rotation, 
it is frequently contaminated by outflow contributions. 
Therefore, CH$_3$CN, SO$_2$, and SiO alone are therefore insufficient to unambiguously probe the disk structures.

\item For I16484$-$4603 and I17008$-$4040, 
double power-law fits to the H$_2$O and SO$2$ rotation curves 
reveal a transition from Keplerian rotation in the inner region 
to infalling-rotating envelope kinematics at larger radii. 
The edge method yields dynamical masses of 46 and 20~$M{\odot}$ 
and disk-envelope transition radii of 318 and 153~au, respectively. 
The confinement of the H$_2$O emission within the transition radii, 
together with the more extended distribution of SO$_2$, 
confirms that H$_2$O is a powerful tracer of the innermost Keplerian disk, 
whereas SO$_2$ predominantly traces the surrounding infalling-rotating envelope.

\item The combination of vibrationally excited H$_2$O and tracers
associated with high-temperature processing of refractory dust (e.g. NaCl, SiS, and SiO)
provides a powerful and relatively unambiguous diagnostic of 
disk-scale rotation in massive protostars. 
However, these lines are intrinsically faint (compared to hot core lines) 
and spatially compact, requiring high sensitivity and angular resolution 
(e.g., $\sim$100 au scales and $\lesssim$2–3 K line sensitivity) 
for robust detection.

\item The high detection rate of disk-like kinematics in our sample suggests that 
hot-disk chemical patterns -- and thus compact rotating disks -- are common 
in massive star-forming regions, 
at least among sources with well-developed rotating envelopes. 
The tentative anti-correlation between disk detections and H30$\alpha$ emission 
further hints that such hot molecular disks may preferentially 
exist at relatively early evolutionary stages, 
prior to significant ionization of the central source.

\end{enumerate}

\begin{acknowledgments}
K.Y. acknowledges supports from the National Natural Science Foundation of China under Grant Number 12503031, the Postdoctoral Fellowship Program of CPSF under Grant Number GZC20252099, the Shanghai Post-doctoral Excellence Program (No. 2024379), the Natural Science Foundation of Shanghai (No. 25ZR1402267) and the Yangyang Development Fund.
Y.Z. acknowledges supports from the Yangyang Development Fund.
N.S. and Z.Z. are supported by a Grant-in-Aid from the Ministry of Education, Culture, Sports, Science, and Technology of Japan (JP20H05844, JP20H05845 and JP25H00676) and a pioneering project in RIKEN (Evolution of Matter in the Universe).
T.L. acknowledges the supports by the National Key R\&D Program of China no. 2022YFA1603100, National Science and Technology Major Project 2024ZD1100601,  National Natural Science Foundation of China (NSFC) through grants no. 12073061 and no. 12122307, the Tianchi Talent Program of Xinjiang Uygur Autonomous Region and the Tianshan Talent Training Program 2024TSYCTD0013.

This paper makes use of the following ALMA data: ADS/JAO.ALMA\#2024.1.01198.S. 
ALMA is a partnership of ESO (representing its member states), NSF (USA) and NINS (Japan), 
together with NRC (Canada), NSTC and ASIAA (Taiwan), and KASI (Republic of Korea), 
in cooperation with the Republic of Chile. 
The Joint ALMA Observatory is operated by ESO, AUI/NRAO and NAOJ. 
The National Radio Astronomy Observatory is a facility of the National Science Foundation 
operated under cooperative agreement by Associated Universities, Inc.
\end{acknowledgments}

\appendix

\section{H30$\alpha$ moment-0 maps} 
\label{sec:app_h30a}

\begin{figure*}[ht!]
\centering
\includegraphics[width=17cm]{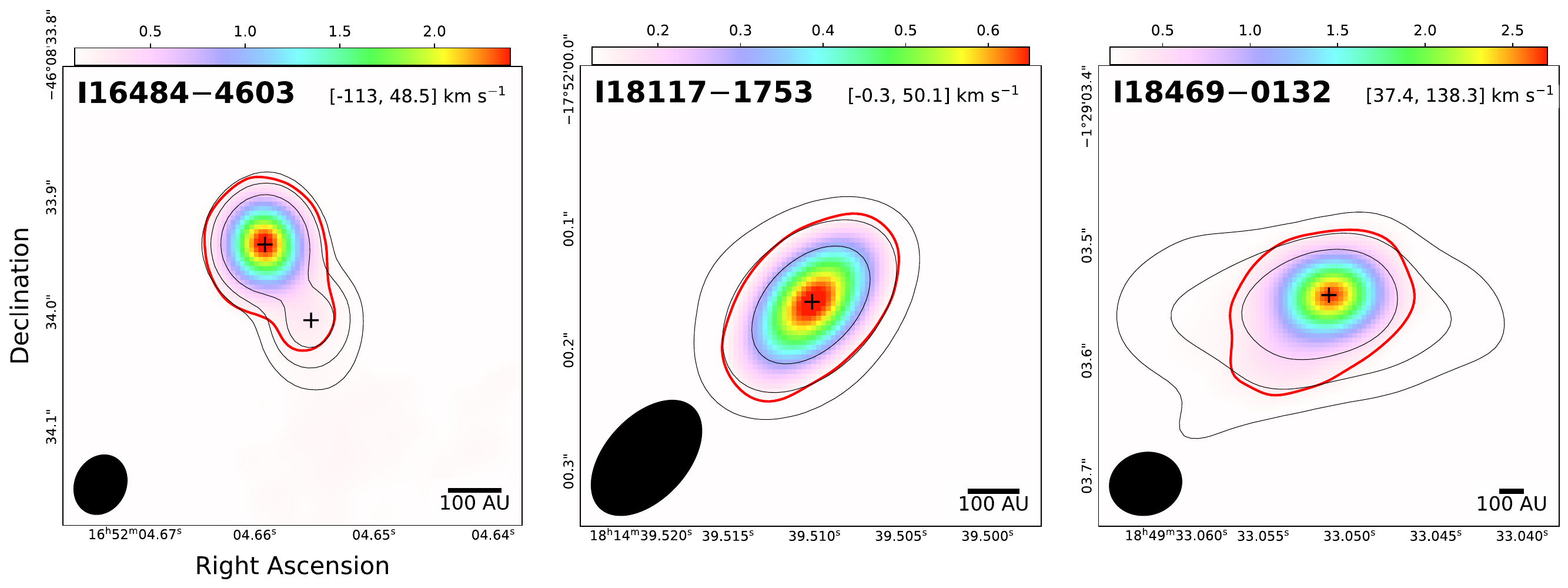}
\caption{
H30$\alpha$ moment-0 maps (color scale) for I16484$-$4603, I18117$-$1753, and I18469$-$0132. The integrated $v_{\rm lsr}$ ranges are labeled in each panel.
The black contours show the continuum emission at levels of 
$10\sigma_{\rm conti} \times 2^{n}$ ($n=4, 5, 6$), while the red contours mark the 
$20\sigma_{\rm area}$ level of the H30$\alpha$ integrated emission, where $\sigma_{\rm area} = \sigma_{\rm chan} \sqrt{\Delta V_{\rm int}\,\delta V_{\rm chan}}$.
The corresponding rms noise levels are listed in Table~\ref{tab:data}. }
\label{fig:h30a}
\end{figure*}

We present the H30$\alpha$ moment-0 maps for I16484$-$4603, I18117$-$1753, and I18469$-$0132
in Figure \ref{fig:h30a}.

\section{Averaged spectra of hot-disk species} 
\label{sec:app_spec}

\begin{figure*}[ht!]
\centering
\includegraphics[width=17cm]{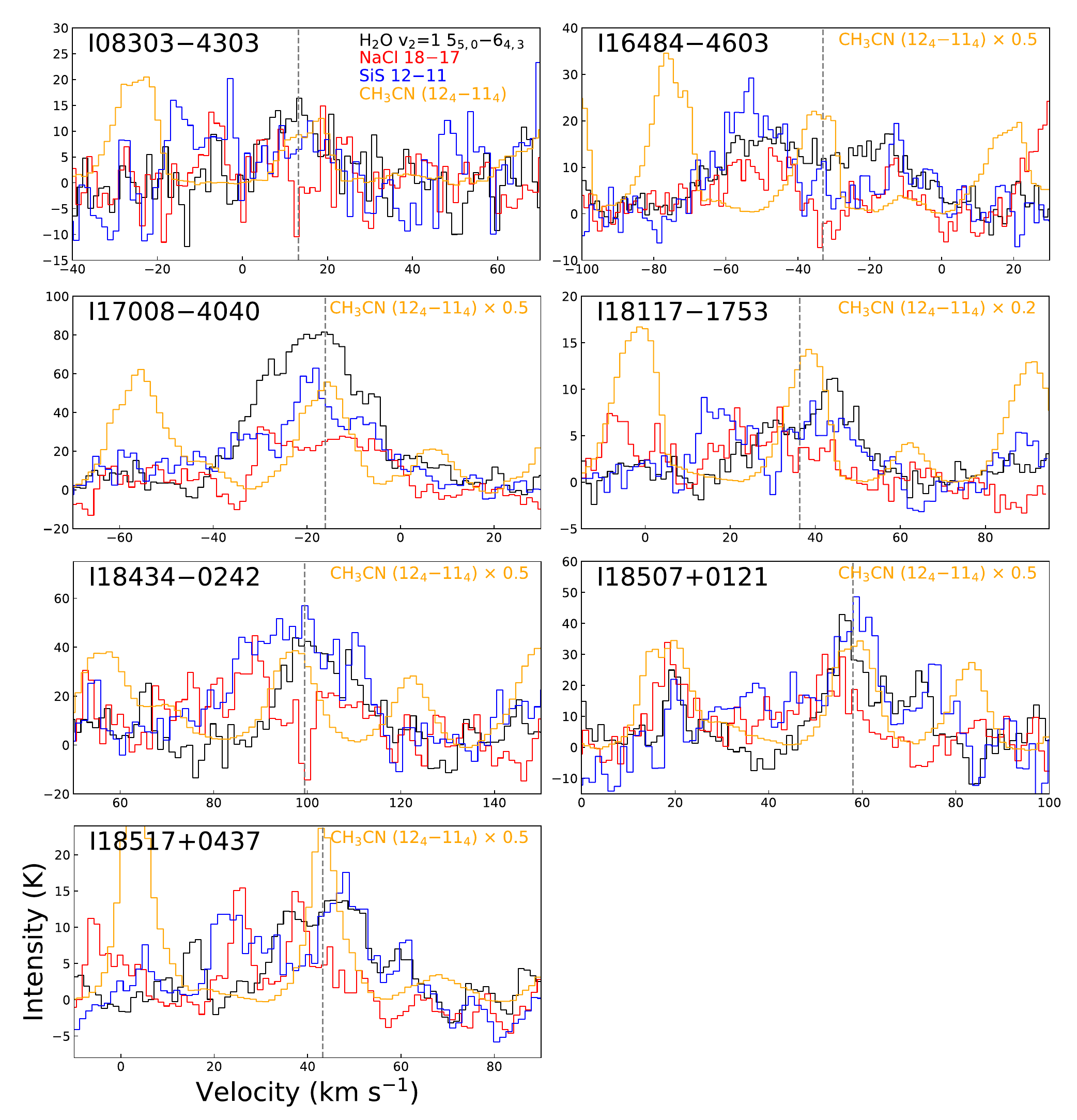}
\caption{Averaged spectra of of H$_2$O ($v_2=1$; $5_{5,0}-6_{4,3}$),
NaCl ($v=0$; $18-17$), SiS ($v=0$; $12-11$) and CH$_3$CN (12$_4$$-$11$_4$) transitions toward the hot-disk sources. 
The H$_2$O, NaCl, and SiS spectra are extracted from regions with integrated emission above $3\sigma$. The CH$_3$CN spectra are extracted from the central 0.3$^{\prime\prime}$ region (about one beam size of the TM2 observation). The vertical dashed lines represent the systemic velocities $v_{\rm sys}$.}
\label{fig:spectra}
\end{figure*}

We present the averaged spectra of of H$_2$O ($v_2=1$; $5_{5,0}-6_{4,3}$),
NaCl ($v=0$; $18-17$) and SiS ($v=0$; $12-11$) transitions for the hot-disk sources 
in Figure \ref{fig:spectra}.

\section{Infrared Spectral Energy Distribution (SED) Fitting}
\label{sec:app_sed}

We construct spectral energy distributions (SEDs) for the ten sources 
from near-infrared to submillimeter wavelengths 
using archival images from {\it Spitzer}, {\it WISE}, {\it SOFIA}, {\it Herschel}, 
and {\it APEX}. 
Photometry is performed using the open-source Python package \texttt{sedcreator} 
\citep{2023ApJ...942....7F}.

Following the standard procedure, we first apply the optimal aperture algorithm 
to the $70~\mu$m image to determine the aperture size. 
If the resulting aperture radius is valid 
(i.e., smaller than the maximum extent of the ALMA continuum emission), 
we adopt this value; 
otherwise, we use a default aperture radius of $20^{\prime\prime}$. 
As in \citet{2023ApJ...942....7F}, we estimate flux uncertainties 
using the fluctuation error method. 
To account for contamination from surrounding cold clump emission, 
we adopt the total background flux as the uncertainty for wavelengths $\ge 100~\mu$m.

We then fit the SEDs of individual sources using \texttt{sedcreator}, 
employing the massive protostellar radiative transfer models of \citet{Zhang2018}. 
These models form a self-consistent grid based on core accretion theory and 
are parameterized by five quantities: 
the initial core mass $M_{\rm c}$, 
the environmental clump mass surface density $\Sigma_{\rm cl}$, 
the protostellar mass $m_{*}$, 
the viewing angle $\theta_{\rm view}$, 
and the foreground extinction $A_{\rm V}$. 
During the fitting, fluxes at wavelengths shorter than $8~\mu$m are 
treated as upper limits (see \citealt[]{2023ApJ...942....7F} for details).
Other parameters, including core radius, current envelope mass, and bolometric luminosity,
are also provided by the fitting.

\begin{figure*}[ht!]
\centering
\includegraphics[width=17cm]{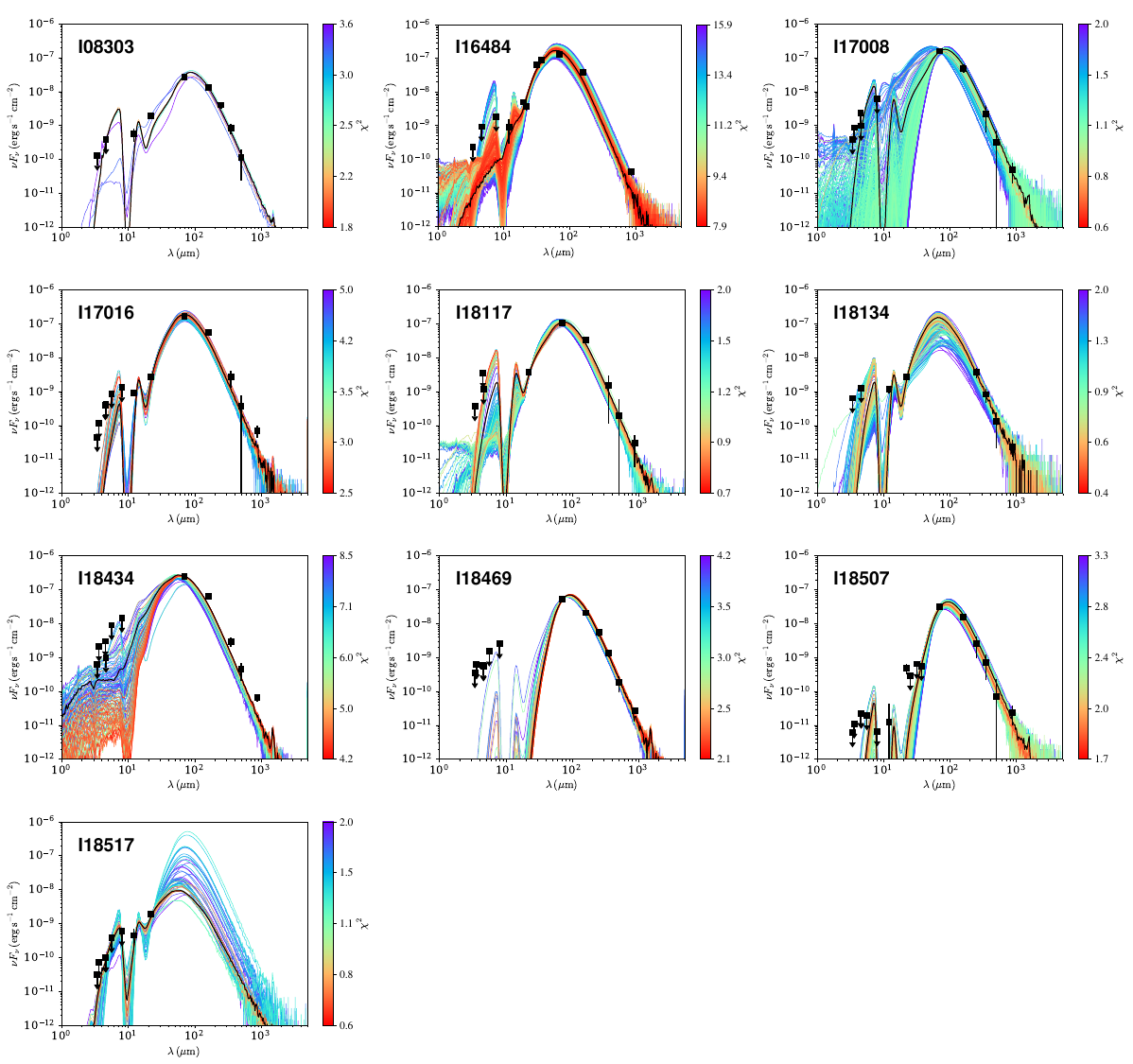}
\caption{Spectral energy distribution (SED) fitting results for the ten sources. 
Observed data points are shown as black symbols with error bars. 
Model SEDs are color-coded by their $\chi^{2}$ values. 
Only models satisfying $R_{\mathrm{core}} < 2R_{\mathrm{aperture}}$ 
and with $\chi^{2}$ between $\chi^{2}_{\mathrm{min}}$ and $\max(2,\,2\chi^{2}_{\mathrm{min}})$ 
(good models) are shown.}
\label{fig:sedfit}
\end{figure*}

The fitting results are shown in Figure~\ref{fig:sedfit}. 
We select ``good models'' with $R_{\rm core}$ smaller than twice the aperture radius and 
with $\chi^{2}$ values between $\chi^{2}_{\rm min}$ and $\max(2,\,2\chi^{2}_{\rm min})$. 
These models are averaged to estimate the protostellar mass $m_{*}$ and 
the intrinsic bolometric luminosity $L_{\rm bol}$ 
(corrected for anisotropic radiation and foreground extinction). 
The derived values of $m_{*}$ and $L_{\rm bol}$ for all sources are listed in Table~\ref{tab:info}.

\bibliography{reference}{}
\bibliographystyle{aasjournalv7}

\newpage

\begin{deluxetable*}{c cc c c ccccc cc cc ll}
    \tabletypesize{\footnotesize}
    \tablecaption{Target Properties and Detection Summary of Disk-Tracing Lines.\label{tab:info}}
    \tablehead{
        \colhead{Source\tablenotemark{a}} & \colhead{R.A.\tablenotemark{b}} & \colhead{Decl.\tablenotemark{b}} & \colhead{Distance\tablenotemark{c}} & \colhead{$V_{\rm sys}$\tablenotemark{d}} & & \multicolumn{3}{c}{Disk-tracer Detection} & & \colhead{SiO\tablenotemark{e}} & \colhead{H30$\alpha$\tablenotemark{f}} & \colhead{$V_{\star}$\tablenotemark{g}} & \colhead{$M_{\star,\mathrm{dyn}}$\tablenotemark{g}} & \colhead{$M_{*,\mathrm{SED}}$\tablenotemark{h}} & \colhead{$L_{\text{bol}}$\tablenotemark{h}}\\
        \cline{7-9}
        \colhead{} & \colhead{(ICRS)} & \colhead{(ICRS)} & \colhead{} & \colhead{} & \colhead{} & \colhead{H$_{2}$O} & \colhead{NaCl} & \colhead{SiS} & \colhead{} & \colhead{} & \colhead{} & \colhead{} & \colhead{} \\
        \colhead{} & \colhead{(hh:mm:ss)} & \colhead{(dd:mm:ss)} & \colhead{(kpc)} & \colhead{(km s$^{-1}$)} & \colhead{} & \colhead{} & \colhead{} & \colhead{} & \colhead{} & \colhead{} & \colhead{} & \colhead{(km s$^{-1}$)} & \colhead{($M_{\odot}$)} & \colhead{($M_{\odot}$)} & \colhead{($10^4~L_{\odot}$)}
    }
    \startdata
    I08303$-$4303 & 08:32:08.34 & $-$43:13:54.0 & 2.35 &    14.3 & & $\surd$ &         &         & & U   &         & 13.2 & 1.6 & $3.6^{+1.2}_{-0.9}$ & $0.8^{+0.2}_{-0.2}$ \\
    I16484$-$4603 & 16:52:04.29 & $-$46:08:30.1 & 2.20 & $-$33.0 & & $\surd$ & $\surd$ & $\surd$ & & D+O & $\surd$ & $-$33.0 & 46 & $20^{+12}_{-8}$     & $13^{+18}_{-8}$ \\
    I17008$-$4040 & 17:04:23.03 & $-$40:44:24.9 & 1.33 & $-$18.0 & & $\surd$ & $\surd$ & $\surd$ & & D+O &         & $-$16.0 & 20 & $15^{+18}_{-8}$     & $8.8^{+27.0}_{-6.6}$ \\
    I17016$-$4124 & 17:05:11.02 & $-$41:29:07.8 & 3.16 & $-$27.1 & &         &         &         & & O   &         &  &  & $37^{+22}_{-14}$    & $48^{+49}_{-24}$ \\
    I18117$-$1753 & 18:14:39.14 & $-$17:52:01.3 & 2.40 &    38.0 & & $\surd$ & $\surd$ & $\surd$ & & D   & $\surd$ & 36.4 & 19 & $17^{+8}_{-6}$    & $10^{+12}_{-6}$\\
    I18134$-$1942 & 18:16:22.12 & $-$19:41:27.0 & 1.25 &    10.6 & &         &         &         & & D   &         &  & & $15^{+18}_{-8}$     & $7.1^{+23.0}_{-5.4}$\\
    I18434$-$0242 & 18:46:03.51 & $-$02:39:26.7 & 5.26 &    97.2 & & $\surd$ & $\surd$ & $\surd$ & & D+O &         & 99.4 & 7.5 & $33^{+16}_{-11}$    & $40^{+21}_{-14}$ \\
    I18469$-$0132 & 18:49:33.15 & $-$01:29:04.2 & 4.91 &    87.0 & &         &         &         & & O   & $\surd$ &  & & $24^{+8}_{-6}$      & $23^{+15}_{-9}$ \\
    I18507$+$0121 & 18:53:18.12 &   +01:25:22.7 & 3.03 &    57.9 & & $\surd$ &         &         & & D   &         & 58.0 & 1.2 & $17^{+18}_{-9}$     & $10^{+29}_{-8}$ \\
    I18517$+$0437 & 18:54:14.13 &   +04:41:46.2 & 1.88 &    43.9 & & $\surd$ & $\surd$ & $\surd$ & & D+O &         & 43.2 & 16 & $15^{+13}_{-7}$     & $5.2^{+14.0}_{-3.8}$ \\
    \enddata
    \tablenotetext{a}{The prefix ``I" in the source names denotes IRAS.}
    \tablenotetext{b}{Phase center coordinates.}
    \tablenotetext{c}{The distances to six sources are obtained from trigonometric parallax measurements: I16484$-$4603 \citep{2015ApJ...805..129K}; I18117$-$1753 \citep{2013Immer}; I18134$-$1942 and I18517$+$0437 \citep{2014Wu}; I18434$-$0242 \citep{2014ApJ...793...72S}; I18507$+$0121 \citep{2023ApJ...949...10M}. The distances to the remaining four sources are kinematic distances taken from \cite{2021MNRAS.505.2801L}.}
    \tablenotetext{d}{Systemic velocities derived from isotopic lines of CH$_{3}$OH and H$_2$CO.}
    \tablenotetext{e}{Detection of SiO: O for tracing outflow, D for tracing disk, U for unclear.}
    \tablenotetext{f}{Detection of H30$\alpha$. The moment-0 maps are presented in Appendix \ref{sec:app_h30a}.}
    \tablenotetext{g}{LSR velocities and dynamical masses of the central stars obtained from the H$_{2}$O PV diagrams, assuming Keplerian rotation and an inclination angle of 90$^\circ$. The results for I16484$-$4603 and I17008$-$4040 are obtained by fitting a double power-law model to the edges of the PV diagrams (see Figure \ref{fig:pv_curves}).}
    \tablenotetext{h}{Protostar mass and bolometric luminosity $L_{\text{bol}}$ from SED fitting (see Appendix \ref{sec:app_sed}).}
    \end{deluxetable*}

\newpage

\begin{deluxetable*}{lccccccccc}
    \tabletypesize{\scriptsize}
    \tablewidth{0pt}
    \tablecaption{Summary of Observations.\label{tab:obs}}
    \tablehead{
        \colhead{Source} & \colhead{Config.} & \colhead{Obs. Date} & \colhead{Ant. \#} & \colhead {$L_\mathrm{min}$ (m)\tablenotemark{a}} & \colhead{$L_\mathrm{max}$ (km)\tablenotemark{b}} & \colhead{$T_\mathrm{int}$ (min)\tablenotemark{c}} & \colhead{Phase Cal.} & \colhead{Bandpass/Flux Cal.}
    }
    \startdata
    \multirow{2}{*}{I08303$-$4303} 
    & TM1 & 2025-07-08 & 42 & 33.7 & 8.8 & 48.6 & \multirow{2}{*}{J0828-3731} & J1107-4449\\
    & TM2 & 2025-03-15 & 47 & 15.1 & 1.4 & 10.6 & & J1037-2934\\
    \hline
    \multirow{2}{*}{I16484$-$4603} 
    & TM1 & 2025-07-08 & 43 & 33.7 & 8.8 & 49.2 & \multirow{2}{*}{J1706-4600} & \multirow{2}{*}{J1617-5848}\\
    & TM2 & 2025-03-21 & 46 & 15.1 & 1.4 & 11.1 & & \\
    \hline
    \multirow{3}{*}{I17008$-$4040} 
    & TM1 & 2025-09-20 & 49 & 33.7 & 11.8& 24.1 & J1720-3552 & \multirow{3}{*}{J1617-5848}\\
    & TM1 & 2025-09-22 & 46 & 33.7 & 9.0 & 24.2 & J1711-3744 & \\
    & TM2 & 2025-03-21 & 46 & 15.1 & 1.4 & 12.6 & J1711-3744 & \\
    \hline
    \multirow{3}{*}{I17016$-$4124} 
    & TM1 & 2025-09-20 & 49 & 33.7 & 11.8& 24.1 & J1720-3552 & \multirow{3}{*}{J1617-5848}\\
    & TM1 & 2025-09-22 & 46 & 33.7 & 9.0 & 24.1 & J1711-3744 & \\
    & TM2 & 2025-03-21 & 46 & 15.1 & 1.4 & 12.7 & J1711-3744 & \\
    \hline
    \multirow{2}{*}{I18117$-$1753} 
    & TM1 & 2025-07-11 & 42 & 33.7 & 8.8 & 46.8 & J1825-1718 & \multirow{2}{*}{J1924-2914}\\
    & TM2 & 2025-03-18 & 45 & 15.1 & 1.4 & 10.6 & J1832-2039 & \\
    \hline
    \multirow{2}{*}{I18134$-$1942} 
    & TM1 & 2025-07-08 & 43 & 33.7 & 8.8 & 46.8 & J1828-2123 & \multirow{2}{*}{J1924-2914}\\
    & TM2 & 2025-03-18 & 44 & 15.1 & 1.4 & 10.6 & J1832-2039 & \\
    \hline
    \multirow{2}{*}{I18434$-$0242} 
    & TM1 & 2025-07-15 & 45 & 59.8 & 11.8& 48.6 & \multirow{2}{*}{J1851+0035} & \multirow{2}{*}{J1924-2914}\\
    & TM2 & 2025-03-18 & 44 & 15.1 & 1.4 & 10.6 & & \\
    \hline
    \multirow{2}{*}{I18469$-$0132} 
    & TM1 & 2025-07-09 & 41 & 113  & 8.8 & 48.9 & \multirow{2}{*}{J1851+0035} & \multirow{2}{*}{J1924-2914}\\
    & TM2 & 2025-03-17 & 44 & 15.1 & 1.1 & 11.1 & & \\
    \hline
    \multirow{2}{*}{I18507$+$0121} 
    & TM1 & 2025-07-08 & 43 & 33.7 & 8.8 & 49.5 & \multirow{2}{*}{J1851+0035} & \multirow{2}{*}{J1924-2914}\\
    & TM2 & 2025-03-17 & 44 & 15.1 & 1.1 & 11.1 & & \\
    \hline
    \multirow{3}{*}{I18517$+$0437} 
    & TM1 & 2025-06-17 & 43 & 92.1 & 8.3 & 25.4 & \multirow{3}{*}{J1851+0035} & \multirow{3}{*}{J1924-2914}\\
    & TM1 & 2025-06-25 & 41 & 33.7 & 6.9 & 25.4 & & \\
    & TM2 & 2025-03-18 & 44 & 15.1 & 1.4 & 11.1 & & \\
    \hline
    \enddata
\tablenotetext{a}{Minimum baseline length.}
\tablenotetext{b}{Maximum baseline length.}
\tablenotetext{c}{On-source integration time.}
\end{deluxetable*}


\newpage

\begin{deluxetable*}{l r lcc c lcc c lc}
    \tablecaption{Summary of Image Data Properties.\label{tab:data}}
    \tablehead{
        \colhead{Source} & \colhead{Robust} & \multicolumn{3}{c}{Low resolution (TM2)} & & \multicolumn{4}{c}{High resolution (TM1+TM2)} \\
        \cline{3-5} \cline{7-12} 
         & & \multicolumn{3}{c}{Continuum} & & \multicolumn{3}{c}{Continuum} & & \multicolumn{2}{c}{Spectral line} \\
        \cline{3-5} \cline{7-9} \cline{11-12}
         & & \colhead{Synthesized Beam} & \colhead{rms} & \colhead{Peak Intensity} & & \colhead{Synthesized Beam} & \colhead{rms} & \colhead{Peak Intensity} & & \colhead{Synthesized Beam\tablenotemark{a}} & \colhead{rms\tablenotemark{b}} \\
         & & \colhead{($^{\prime\prime}~\times~^{\prime\prime},~\pa$)} & ($\mu$Jy beam$^{-1}$) & (mJy beam$^{-1}$) & & \colhead{($^{\prime\prime}~\times~^{\prime\prime},~\pa$)} & ($\mu$Jy beam$^{-1}$) & (mJy beam$^{-1}$) & & \colhead{($^{\prime\prime}~\times~^{\prime\prime},~\pa$)} & ($\mJybeam$)
    }
    \startdata
    I08303$-$4303 & 0.5 & $0\farcs40\times0\farcs34,~1^\circ$   & 60 & 28.3 & & $0\farcs052\times0\farcs044,~43^\circ$  & 20 & 7.33 & & $0\farcs049\times0\farcs042,~42^\circ$ & 0.6 \\
    I16484$-$4603 & 0.5 & $0\farcs29\times0\farcs28,~42^\circ$  & 60 & 80.0 & & $0\farcs056\times0\farcs047,~23^\circ$  & 20 & 50.4 & & $0\farcs053\times0\farcs044,~26^\circ$ & 0.6 \\
    I17008$-$4040 & 0.5 & $0\farcs29\times0\farcs28,~71^\circ$  & 90 & 211  & & $0\farcs050\times0\farcs041,~65^\circ$  & 30 & 46.7 & & $0\farcs047\times0\farcs039,~66^\circ$ & 0.6 \\
    I17016$-$4124 & 0.5 & $0\farcs28\times0\farcs28,~65^\circ$  & 90 & 349  & & $0\farcs050\times0\farcs041,~63^\circ$  & 30 & 38.4 & & $0\farcs047\times0\farcs038,~64^\circ$ & 0.6 \\
    I18117$-$1753$\tablenotemark{c}$ & 0.5 & $0\farcs36\times0\farcs28,~-82^\circ$ & 90 & 37.6 & & $0\farcs113\times0\farcs069,~-45^\circ$ & 30 & 37.2 & & $0\farcs111\times0\farcs065,~-42^\circ$ & 0.6 \\
               & $-$0.5 &                                       &    &  & & $0\farcs061\times0\farcs042,~76^\circ$  & 15 & 34.2 & &     \\
    I18134$-$1942 & 0.5 & $0\farcs35\times0\farcs27,~-81^\circ$ & 60 & 39.6 & & $0\farcs082\times0\farcs053,~86^\circ$  & 20 & 22.1 & & $0\farcs073\times0\farcs050,~83^\circ$  & 0.6 \\
    I18434$-$0242 & 0.5 & $0\farcs35\times0\farcs28,~-76^\circ$ & 20 & 93.3 & & $0\farcs053\times0\farcs046,~-57^\circ$ & 40 & 16.2 & & $0\farcs052\times0\farcs045,~-55^\circ$ & 0.6 \\
    I18469$-$0132 & 0.5 & $0\farcs49\times0\farcs33,~-82^\circ$ & 60 & 128  & & $0\farcs066\times0\farcs054,~-82^\circ$ & 20 & 39.5 & & $0\farcs063\times0\farcs053,~-86^\circ$ & 0.6 \\
    I18507$+$0121 & 0.5 & $0\farcs46\times0\farcs34,~-84^\circ$ & 75 & 187  & & $0\farcs064\times0\farcs052,~44^\circ$  & 25 & 19.6 & & $0\farcs062\times0\farcs049,~41^\circ$  & 0.6 \\
    I18517$+$0437 & 0.5 & $0\farcs35\times0\farcs31,~68^\circ$  & 72 & 51.3 & & $0\farcs105\times0\farcs073,~-65^\circ$ & 24 & 18.2 & & $0\farcs100\times0\farcs071,~-65^\circ$ & 0.6 \\
    \enddata
    \tablenotetext{a}{Synthesized beams of image cubes
    at representative frequency of 232.6867 GHz (the rest frequency of the H$_{2}$O $v_{2}=1$ $5_{5,0}-6_{4,3}$ transition), as the beam shape
    slightly changes with the frequency.}
    \tablenotetext{b}{Typical noise levels for the spectral windows per channel, as the noise level slightly changes with the frequency.
See the image captions of the presented lines for more information.}
    \tablenotetext{c}{I18117$-$1753 was imaged with a Briggs robust $-$0.5 to achieve a higher resolution continuum map.}
\end{deluxetable*}

\newpage


\begin{table*}
\centering
\caption{Parameters of the spectral lines\tablenotemark{a}.}  \label{tab:lines}
\begin{tabular}{l l l l l}
\hline\hline
Molecule & Transition & Rest frequency & $E_{u}/k$ & $A_{ul}$ \\
 &  & (MHz) & (K) & (s$^{-1}$) \\
\hline
 CH$_{3}$CN & $12_{4}-11_{4}$ & 220679.2869 & 183.1 & 8.2$\times$10$^{-4}$ \\
 SO$_{2}$ & $28_{3,25}-28_{2,26}$ & 234187.0540 & 403.0 & 1.4$\times$10$^{-4}$ \\
 SiO & $5-4$ & 217104.9800 & 31.3 & 5.2$\times$10$^{-4}$ \\
 H$_{2}$O & $5_{5,0}-6_{4,3}$ ($v_{2}=1$) & 232686.7000 & 3461.9 & 4.8$\times$10$^{-6}$ \\
 NaCl & $18-17$ ($v = 0$) & 234251.8300 & 106.8 & 5.9$\times$10$^{-3}$ \\
 NaCl & $18-17$ ($v = 1$) & 232509.9500 & 625.7 & 5.8$\times$10$^{-3}$ \\
 NaCl & $17-16$ ($v = 1$) & 219614.9220 & 614.5 & 4.9$\times$10$^{-3}$ \\
 NaCl & $17-16$ ($v = 2$) & 217980.2149 & 1128.4 & 4.9$\times$10$^{-3}$ \\
 SiS & $12-11$ ($v = 0$) & 217817.6630 & 68.0 & 1.7$\times$10$^{-4}$ \\
 SiS & $13-12$ ($v = 1$) & 234812.9678 & 1150.0 & 2.2$\times$10$^{-4}$ \\
 PN & $5-4$ ($v = 0$) & 234935.6630 & 33.8 & 5.2$\times$10$^{-4}$ \\
 PN & $5-4$ ($v = 1$) & 233271.8000 & 1937.3 & 5.0$\times$10$^{-4}$ \\
\hline\hline
\end{tabular}
 \tablenotetext{a}{The spectroscopic parameters are obtained from the CDMS database \citep{2001A&A...370L..49M,2005JMoSt.742..215M,2016JMoSp.327...95E} and JPL catalogues \citep{1998JQSRT..60..883P}.}
\end{table*}

\begin{rotatetable}
\begin{deluxetable*}{c c ccccccc ccccc cc}
    \tablecaption{Peak intensity of the disk-tracing lines. \label{tab:peak}}
    \tablehead{
        \colhead{Source} & \colhead{H$_2$O} & & \multicolumn{5}{c}{NaCl} & & \multicolumn{2}{c}{SiS} & & H(30)$\alpha$ & Continuum \\
        \cline{4-8} \cline{10-11} 
        \colhead{} & \colhead{5$_{5,0}$$-$6$_{4,3}$} & & \multicolumn{2}{c}{18$-$17} & & \multicolumn{2}{c}{17$-$16} & & \colhead{12$-$11} & \colhead{13$-$12} & &  \\
        \cline{4-5} \cline{7-8} 
        \colhead{} & \colhead{($v_{2}$=1)} & & \colhead{($v$ = 0)} & \colhead{($v$ = 1)} & & \colhead{($v$ = 1)} & \colhead{($v$ = 2)} & & \colhead{($v$ = 0)} & \colhead{($v$ = 1)} &  &  &  \\
        \colhead{} & \colhead{(K)} & & \colhead{(K)} & \colhead{(K)} & & \colhead{(K)} & \colhead{(K)} & & \colhead{(K)} & \colhead{(K)} & \colhead{} & \colhead{(K)} & \colhead{(K)}
    }
    \startdata
    I08303$-$4303 & 18.6 & &  --  &  --  & &  --  &  --  & &  --  &  --  & & -- & 24.3 \\ 
    I16484$-$4603 & 37.1 & & 26.1 & 20.9 & & 24.1 & 17.5 & & 30.9 & 20.3 & & 42.3 & 185 \\ 
    I17008$-$4040 & 124  & & 50.0 & 54.5 & & 46.2 & 51.7 & & 59.7 & 52.2 & & -- & 596 \\ 
    I18117$-$1753 & 21.3 & & 11.9 & 8.75 & & 13.8 & 6.65 & & 11.1 &  --  & & 39.1 & 113 \\ 
    I18434$-$0242 & 45.6 & & 51.0 & 33.7 & & 51.8 & 16.6 & & 59.8 & 43.0 & & -- & 166 \\ 
    I18507$+$0121 & 35.4 & &  --  &  --  & &  --  &  --  & &  --  &  --  & & -- & 146 \\ 
    I18517$+$0437 & 22.7 & & 18.8 &  --  & &  --  &  --  & & 25.3 & 8.85 & & -- & 56.7 \\ 
    \enddata
\end{deluxetable*}
\end{rotatetable}

\begin{deluxetable*}{cc c ccc c ccc}
    \tabletypesize{\normalsize}
    \tablecaption{Double power-law fitting to the rotation curves in I16484$-$4603 and I17008$-$4040.}\label{tab:pv_curve}
    \tablehead{
        \colhead{Source} & \colhead{Method} & \colhead{} & \multicolumn{3}{c}{Double Power Law} & \colhead{} & \multicolumn{3}{c}{Inner Power Law Fixed ($p_{\rm in}$=0.5)} \\
    \cline{4-6} \cline{8-10}
        \colhead{} & \colhead{} & \colhead{} & \colhead{$p_{\rm in}$\tablenotemark{a}} & \colhead{$p_{\rm out}$\tablenotemark{b}} & \colhead{$r_{\rm b}$\tablenotemark{c}} & \colhead{} & \colhead{$p_{\rm out}$\tablenotemark{b}} & \colhead{$r_{\rm b}$\tablenotemark{c}} & \colhead{$M_\star$\tablenotemark{d}} \\
        \colhead{} & \colhead{} & \colhead{} & \colhead{} & \colhead{} & \colhead{(au)} & \colhead{} & \colhead{} & \colhead{(au)} & \colhead{($M_\odot$)}
    }
    \startdata
    \multirow{2}{*}{I16484$-$4603} 
    & edge  &  & 0.42$\pm$0.03 & 1.27$\pm$0.05 & 318$\pm$34 &  & 1.28$\pm$0.04 & 391$\pm$36 & 46$\pm$3 \\
    & ridge &  & 0.57$\pm$0.06 & 1.00$\pm$0.08 & 111$\pm$32 &  & 1.00$\pm$0.06 & 98$\pm$23 & 33$\pm$1 \\
    \hline
    \multirow{2}{*}{I17008$-$4040} 
    & edge  &  & 0.58$\pm$0.03 & 0.90$\pm$0.03 & 153$\pm$15 &  & 0.90$\pm$0.02 & 239$\pm$17 & 20$\pm$1 \\
    & ridge &  & 0.58$\pm$0.03 & 1.05$\pm$0.03 & 69$\pm$13  &  & 1.05$\pm$0.03 & 61$\pm$15 & 6.3$\pm$0.6 \\
    \hline
    \enddata
\tablenotetext{a}{Index of the inner power law ($v_\mathrm{rot}\propto r^{-p_\mathrm{in}}$).}
\tablenotetext{b}{Index of the outer power law ($v_\mathrm{rot}\propto r^{-p_\mathrm{out}}$).}
\tablenotetext{c}{Breaking radius between the inner and outer power-laws.}
\tablenotetext{d}{Central mass derived from inner Keplerian power-law with $v_\mathrm{rot}\propto r^{-0.5}$.}
\end{deluxetable*}

\newpage

\newpage

\begin{figure*}[ht!]
 \begin{center}
\includegraphics[width=\textwidth]{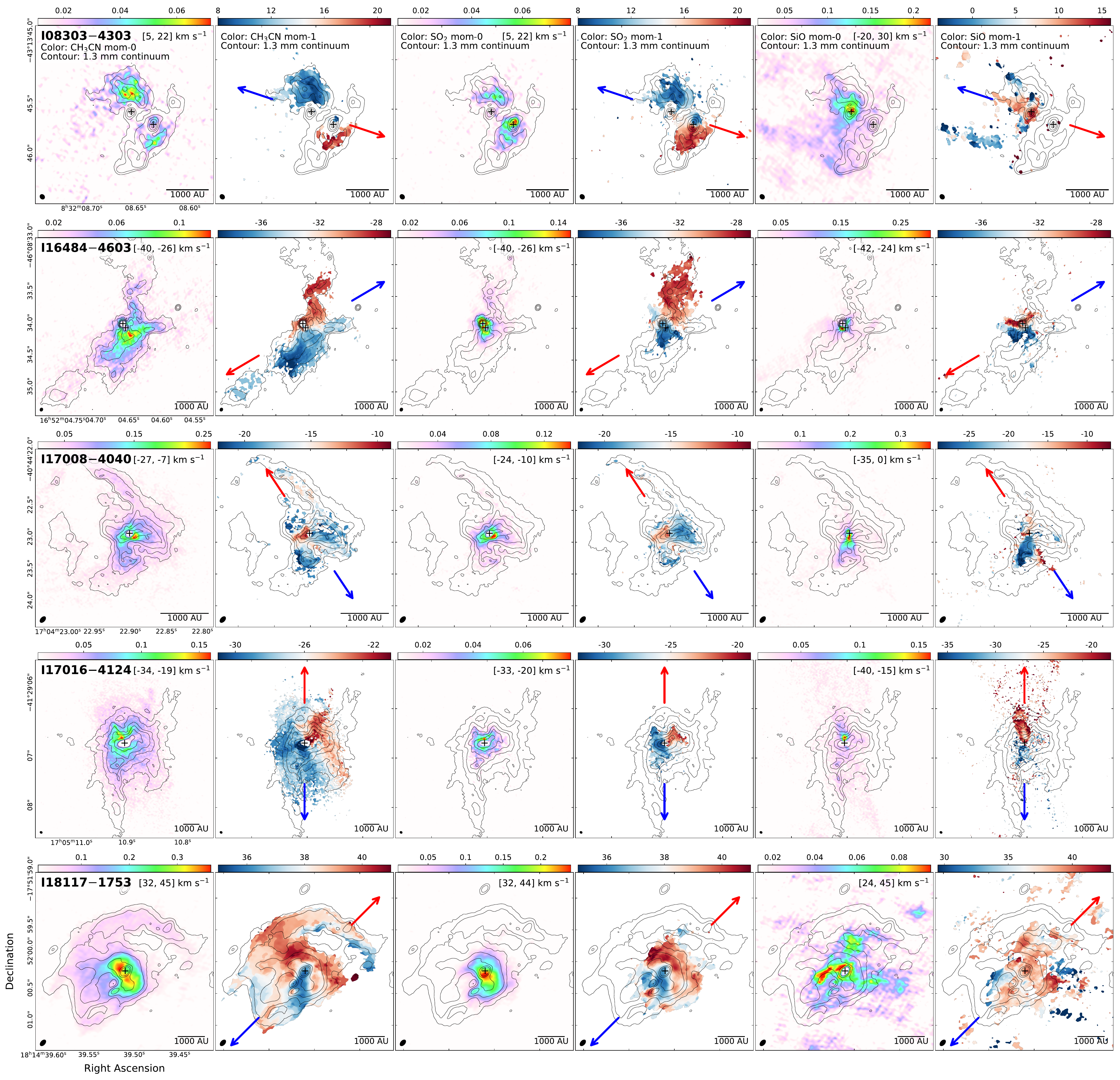}
\caption{Moment 0 and moment 1 maps of CH$_{3}$CN ($12_{4}-11_{4}$), SO$_{2}$ ($28_{3,25}-28_{2,26}$), and SiO ($5-4$) toward our sample. 
Contours show the high-resolution 1.3 mm continuum emission, 
with the same levels as in Figure~\ref{fig:sample_conti}.
The red and blue arrows indicate the outflow directions.
The color bars show the intensity range in Jy beam$^{-1}$ km s$^{-1}$ for the moment-0 maps 
and the velocity range in $\kms$ for the moment-1 maps for the three lines in each source.}
\label{fig:sample_lines}
\end{center}
\end{figure*}

\newpage

\begin{figure*}[ht!]
\addtocounter{figure}{-1}
 \begin{center}
\includegraphics[width=\textwidth]{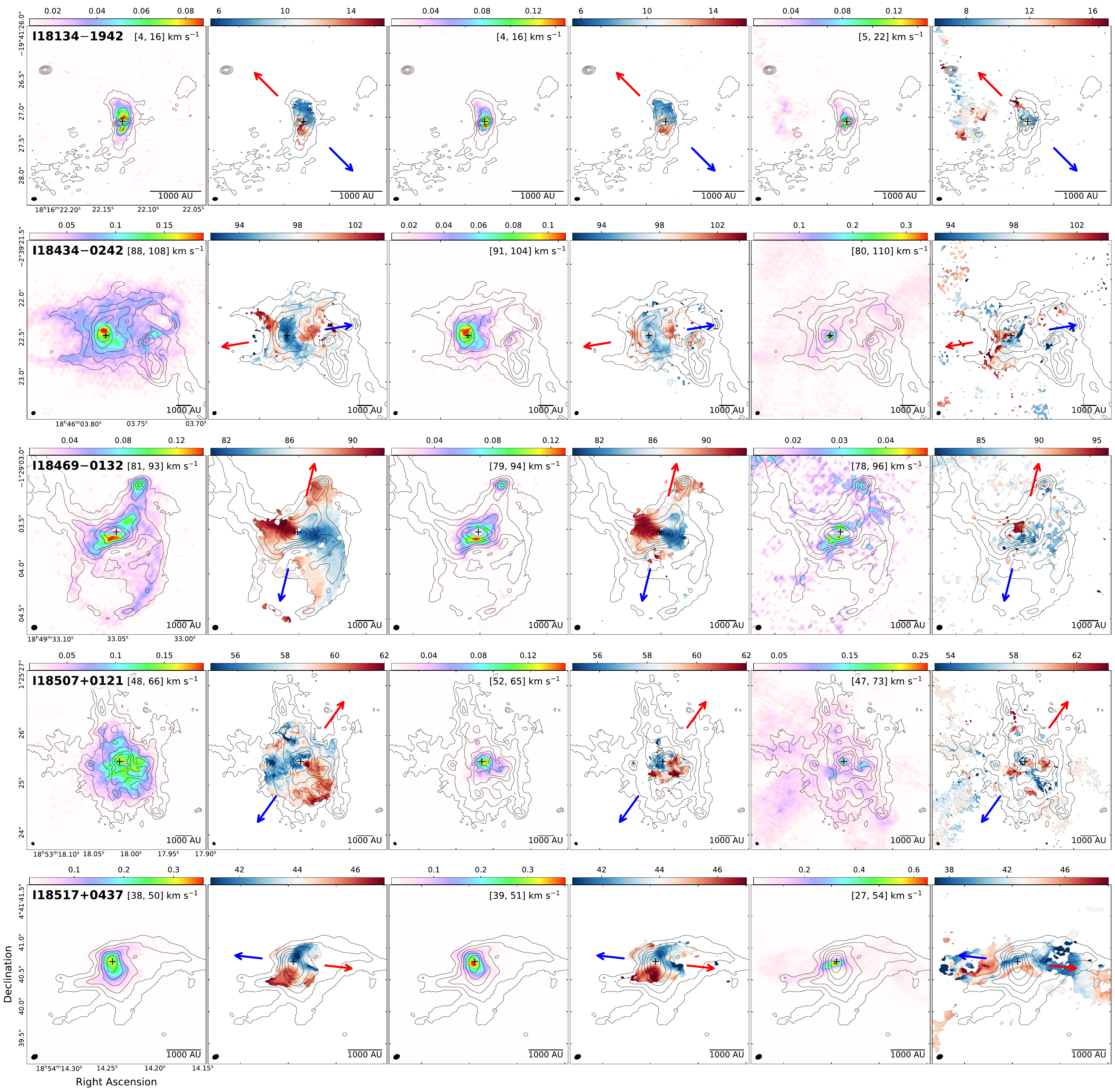}
\caption{(Continued.)}
\end{center}
\end{figure*}

\newpage

\newpage

\end{document}